\documentclass[aps,pra,reprint,showpacs,nofootinbib,superscriptaddress]{revtex4-1}

\expandafter\let\csname equation*\endcsname\relax

\expandafter\let\csname endequation*\endcsname\relax

\usepackage{graphicx, comment}
\usepackage{amsfonts}
\usepackage{amsmath,amssymb}
\usepackage{bm}
\usepackage{color}
\usepackage{epsfig}
\usepackage{verbatim}
\usepackage{lineno}
\usepackage[thicklines]{cancel}
\usepackage{hyperref}
\usepackage{multirow}
\usepackage[table,xcdraw]{xcolor}
\usepackage{tabularx}
\usepackage{wrapfig}
\usepackage[font=small,
  justification=justified,
  format=plain]{caption} 
\usepackage{subfig}
\usepackage{url}   
\usepackage{xcolor}
\usepackage{listings}
\usepackage{ulem}
\usepackage[T1]{fontenc}
\usepackage{algorithm,algorithmic}
\usepackage{enumerate}
\usepackage{siunitx}
\usepackage{booktabs}
\usepackage{hyperref}
\usepackage{cancel}
\hypersetup{
     colorlinks   = true,
     citecolor    = blue
}

\newcommand{\fix}[1]{{\color{black}#1}}
\definecolor{red}{rgb}{1,0.,0}
\newcommand{\expect}[1]{\langle {#1} \rangle}
\def\be{\begin{equation}}
\def\ee{\end{equation}}
\def\bea{\begin{eqnarray}}
\def\eea{\end{eqnarray}}

\begin{document}

\title{Scattering in the Ising Model with the Quantum Lanczos Algorithm }

\author{K\"ubra Yeter-Aydeniz }
\email{yeteraydenik@ornl.gov}
\address{Computational Sciences and Engineering Division, Oak Ridge National Laboratory,
  Oak Ridge, TN 37831, USA}
\address{Physics Division, Oak Ridge National Laboratory,
  Oak Ridge, TN 37831, USA}

\author{George Siopsis}
\email{siopsis@tennessee.edu}
\address{Department of Physics and Astronomy,  The University of Tennessee, Knoxville, TN 37996-1200, USA}

\author{Raphael C.\ Pooser}
\email{pooserrc@ornl.gov}
\address{Computational Sciences and Engineering Division, Oak Ridge National Laboratory,
  Oak Ridge, TN 37831, USA}
\address{Physics Division, Oak Ridge National Laboratory,
  Oak Ridge, TN 37831, USA}
\address{Department of Physics and Astronomy,  The University of Tennessee, Knoxville, TN 37996-1200, USA}

%


\date{\today}
\begin{abstract}

Time evolution and scattering simulation in phenomenological models are of great interest for testing and validating the potential for near-term quantum computers to simulate quantum field theories. Here, we simulate one-particle propagation and two-particle scattering in the one-dimensional transverse Ising model for 3 and 4 spatial sites with periodic boundary conditions on a quantum computer. We use the quantum Lanczos algorithm to obtain all energy levels and corresponding eigenstates of the system. We simplify the quantum computation by taking advantage of the symmetries of the system. These results enable us to compute one- and two-particle transition amplitudes, particle numbers for spatial sites, and the transverse magnetization as functions of time. The quantum circuits were executed on \fix{various} IBM Q superconducting hardware. The experimental results
are in very good agreement with the values obtained using exact diagonalization. {\footnote{This manuscript has been authored by UT-Battelle, LLC, under Contract No. DE-AC0500OR22725 with the U.S. Department of Energy. The United States Government retains and the publisher, by accepting the article for publication, acknowledges that the United States Government retains a non-exclusive, paid-up, irrevocable, world-wide license to publish or reproduce the published form of this manuscript, or allow others to do so, for the United States Government purposes. The Department of Energy will provide public access to these results of federally sponsored research in accordance with the DOE Public Access Plan.}}     
\end{abstract}

\maketitle
%
\vspace{2pc}

%
%
%

\section{Introduction}
The Ising model is a quintessential spin system within which one can simulate and study many-body interactions. The model allows for simulating spin-spin physics and the calculation of properties such as magnetization and spin-frustration. \fix{For instance, the Ising model is of critical importance in the study of high $T_c$ superconductors since it allows one to study the electrical transport properties near a quantum critical point (Ising-nematic) which helps one understand strong electronic interactions in these systems \cite{Wang2019}. To obtain information about the non-equilibrium dynamics of isolated many-body systems, the time evolution of the transverse magnetization as well as the entanglement entropy of the evolved states have been evaluated to study the domain wall melting in the ferromagnetic phase of transverse Ising chains \cite{Eisler2016}. The Ising model} also serves as a useful arena for the study of more complex quantum field theories on a lattice. For example, scattering in a spin system on a lattice holds many parallels with scattering between particles in high energy physics experiments~\cite{Gustafson2019_1,Gustafson2019_2,Kim2011,Lamm2018,Smith2019}. 

\fix{In a different perspective, the Ising model itself is used as a generic quantum computer model for the adiabatic quantum computers and quantum annealers~\cite{Albash2021}. Therefore, the dynamics and the correlations of the quantum entanglement between large number of spins are of interest related to the operation of quantum annealers~\cite{Navez2017}.}

Computing scattering amplitudes, transition rates, and other physical quantities involving quantum fields are hard tasks for classical computers. Quantum computers promise exponential speedup, however the approach with quantum simulators often revolves around the computation of real-time evolution based on Trotterization which is of limited utility on NISQ (noisy intermediate-scale quantum~\cite{Preskill2018})  hardware~\cite{klco_quantum-classical_2018}. Previous studies have simulated real-time dynamics of interactions~\cite{Gustafson2019_1,Kim2011,Lamm2018,Smith2019} and evolution of disordered Hamiltonians~\cite{Alexandru2020} with this method. In this type of simulation, the number of gates grows linearly with the system size and the number of Trotter steps. Therefore, the noise in the system grows as the system size grows.

\fix{As an alternative perspective for NISQ devices, the variational quantum simulation of real time, imaginary time, and generalized time evolution of quantum systems have also been studied~\cite{Yuan2019Quantum, Endo2020PRL}. The introduction of variational quantum simulation for studying real-time dynamics of quantum systems and comparison to Trotterization method was first done in~\cite{Li2017} where it was claimed to provide an improvement over the Trotterization method. Specifically in Ref.~\cite{Endo2020PRL}, the authors introduced a variational quantum simulation of open system dynamics and numerically tested the algorithm with a 6-qubit 2D transverse-field Ising model under dissipation. A comprehensive review of these variational algorithms can  be found in~\cite{Endo2020review}. }

Others have simulated the Ising model both variationally~\cite{Seki2020} and via direct diagonalization within the quantum circuit~\cite{Cervera-Lierta2018}. 

Here, we use the Quantum Lanczos (QLanczos) algorithm~\cite{Motta2019} to calculate transition probabilities and scattering amplitudes in the one-dimensional transverse Ising model with periodic boundary conditions. We use the quantum imaginary-time evolution algorithm (QITE) to provide a basis for the Hilbert (Krylov) space employed in the QLanczos algorithm. 
We tune the QITE step size, and thus the total noise in the circuit, by using a hybrid quantum-classical approach to the algorithm.
Using this technique we also compute occupation numbers and the transverse magnetization.

\fix{Extending the results of this study for the use of Ising model as a generic model would be an interesting research but we will leave this as a future study.}

\fix{The hybrid quantum-classical version of the imaginary-time evolution was first proposed in \cite{McArdle2019} where the non-unitary imaginary-time evolution operator was approximated by a parameterized \textit{Ansatz} state, and the parameters to obtain the ground state were found using a variational method. The QITE algorithm proposed in \cite{Motta2019} has certain advantages, because it does not require costly optimization or ancilla qubits. When it comes to its implementation on NISQ devices, it has disadvantages over the method of \cite{McArdle2019} because of increasing circuit depth at each QITE step which raises the impact of noise from short coherence time, cross-talk between qubits, etc. Recent efforts have sought to economize the circuit depth in the QITE algorithm \cite{Yeter2019, Nishi2020, Gomes2020} to reduce the impact of these noise sources.}
In \cite{Yeter2019}, we employed a method that simplified the quantum circuit needed for the unitary updates of the QITE step, thereby reducing the gate depth and noise. Here, we follow a slightly different approach for quantum circuit simplification.

For $N_s=3,4$ spatial sites in the Ising spin chain with periodic boundary conditions, we used the QLanczos algorithm to compute the eigenvalues and eigenstates of the system so that transition probability, occupation number, and transverse magnetization could be calculated. We computed  energy expectation values as functions of imaginary time on the IBM Q 5-qubit Yorktown device. These expectation values were obtained using QITE, and were subsequently fed to the QLanczos algorithm. We benchmarked these results against exact calculations, and obtained good agreement when error mitigation was employed.

\fix{Our discussion is organized as follows. In Section \ref{sec:prelim}, we introduce the model and the physical quantities to be computed. We discuss the Hilbert space and the simplifications afforded by symmetry. In Section \ref{sec:algo}, we discuss the QITE and QLanczos algorithms, and the details of our hybrid classical-quantum implementation. In Section \ref{sec:QProgram}, we discuss the implementation of our quantum algorithm including error mitigation. In Section \ref{sec:Results}, we discuss our results. Finally, in Section \ref{sec:Conclusion}, we summarize our conclusions.
}

\section{Preliminaries}
\label{sec:prelim}

\fix{In this Section, we introduce the Ising model we used in our work and define the physical quantities we computed. We also discuss details of the Hilbert space and the simplifications one can take advantage of due to symmetry.}

\subsection{The model}\label{sec:mod}
The Ising model Hamiltonian with periodic boundary conditions (PBC) can be written as
\be
H =-J \sum_{i\in \mathbb{Z}_{N_s}} X_i X_{i+1}-h_T\sum_{i\in\mathbb{Z}_{N_s}} Z_i~, \label{PBCHam}
\ee
where $X_i, Y_i, Z_i$ are the Pauli matrices at the $i$th site, $i=0,1,\dots,N_s-1$, $N_s$ is the number of spatial sites, $J$ is the nearest-neighbor coupling strength, and $h_T$ is the transverse magnetic field. We impose periodic boundary conditions by identifying \fix{$X_{N_s} = X_0$}.
At each site, we place a qubit on which the Pauli matrices act, and define the occupation number of the $i$th site by $n_i = \frac{\mathbb{I} - Z_i}{2}$ with corresponding eigenstates $|n_i\rangle$, where $n_i=0,1$ ($|0\rangle$ ($|1\rangle$) denotes an unoccupied (occupied) site). A vector in the computational basis $|x\rangle$ ($x=0,1,\dots, 2^{N_s}-1$) is specified by the sites which are occupied corresponding to the digits of $x$ equal to 1 (e.g., for $N_s=4$, the state $|0000\rangle$ has no particles, whereas $|0101\rangle$ consists of two particles at sites 1 and 3).
\subsection{Unitary Time Evolution}
To study the time evolution of the system, we prepare it in the initial state $|\text{initial}\rangle$, evolve it for time $t$ with the evolution operator $\,\mathcal{U}(t)=e^{-iHt}$, and then measure it, thus projecting it onto a state $|\text{final}\rangle$. This process leads to the quantum computation of the transition probability
\be\label{eq:2}
P_{fi} (t) =|\mathcal{A}_{fi}(t)|^2~, \ \ \ \ \mathcal{A}_{fi} (t) =\langle \text{final} \, |\, \mathcal{U}(t) | \text{initial} \rangle~.
\ee
In particular, in this work we study single-particle propagation and two-particle scattering. In both cases, we prepare the system in the computational basis state $|\text{initial}\rangle=|x_{\text{in}}\rangle$. For single-particle propagation, $x_{\text{in}}$ contains a single digit equal to 1, whereas for two-particle scattering, it contains two digits equal to 1. At the end of the quantum computation, the measurement projects the system onto a different computational basis state $|\text{final}\rangle = |x_{\text{fin}}\rangle$. Being in the computational basis, both initial and final states are easy to construct. However, the unitary $\mathcal{U}(t)$ is difficult to implement. We use the QLanczos algorithm to accomplish this, which is based on the quantum imaginary-time evolution (QITE) algorithm~\cite{Motta2019}. 

To calculate the transition probabilities \eqref{eq:2}, we employ a hybrid quantum-classical algorithm to solve the eigenvalue problem of the Hamiltonian \eqref{PBCHam},
\be\label{eq:3}
H|\psi_I\rangle = E_I |\psi_I\rangle \ , \ \ I=0,1,\dots, 2^{N_s}-1~.
\ee
The unitary evolution operator is expressed in terms of the eigenvalues and eigenstates of the Hamiltonian \eqref{PBCHam} as
\be\label{eq:4} \mathcal{U} (t) = \sum_{I=0}^{2^{N_s}-1} e^{-iE_It} |\psi_I\rangle \langle \psi_I |~. \ee
Let $\bm{t}$ be the unitary transformation from the eigenstates of $H$ to the computational basis. Its matrix elements are
\be\label{eq:5} t_{Ix} = \langle \psi_I |x\rangle~. \ee
All components of the eigenstates $|\psi_I\rangle$ are real, therefore, $t_{Ix} \in \mathbb{R}$. This will simplify the computation of the components of the eigenstates.

Scattering data can be expressed in terms of transition amplitudes between an initial and a final state, both members of the computational basis, $|x_{\text{in}}\rangle$ and $|x_{\text{fin}}\rangle$, respectively. A transition amplitude over time $t$,
\be \mathcal{A}_{fi} (t) \equiv \langle x_{\text{fin}} | \mathcal{U} (t) | x_{\text{in}} \rangle \ee
can be calculated classically using the matrix $\bm{t}$ (eq.\ \eqref{eq:5}. We obtain
\be
\mathcal{A}_{fi} (t) = \sum_{I=0}^{2^{N_s}-1} t_{Ix_{\text{in}}} t_{Ix_{\text{fin}}} e^{-iE_It}~.\label{eq:trans}
\ee
\fix{It should be noted that, while this calculation leads to more accurate results for NISQ devices, as we will demonstrate, for a large number of qubits, it may be more efficient to use other approaches, such as Trotterization on the evolution unitary $\mathcal{U} (t)$.}

The time evolution of the occupation number for the $i$th site ($i=1,\dots,N_s$) can be calculated using the expression \eqref{eq:4} of the evolution operator. We obtain the average in the state $|x\rangle$ at time $t$,
\be
\langle x| n_i (t)|x\rangle= \sum_{I,J,y=0}^{2^{N_s}-1} y_i t_{Ix} t_{Jx} t_{Iy} t_{Jy} e^{i(E_J-E_I)t}~, \label{eq:ni}
\ee
where $y_i$ is the $i$th digit in the binary expansion of $y$.
We deduce the transverse magnetization as
\be
\langle m_z(t) \rangle \equiv \frac{1}{N_s}\sum_{i=0}^{N_s-1} \langle Z_{i}(t)\rangle = 1 - \frac{2}{N_s}\sum_{i=0}^{N_s-1} \langle n_{i}(t)\rangle~.
\label{eq:mz}
\ee
One can also simulate the thermal evolution of the system \cite{Cervera-Lierta2018} by computing the ensemble average of any operator $\mathcal{O}$ at finite temperature, $T$,
\be
\langle \mathcal{O}(\beta)\rangle= \frac{1}{\mathcal{Z}}\sum_{I=0}^{2^{N_s}-1} e^{\beta E_I}\langle{\psi_I}|\mathcal{O}|\psi_I\rangle~,
\ee
where $\beta=\frac{1}{k_B T}$, $k_B$ is the Boltzmann constant, and $\mathcal{Z}=\sum_I e^{-\beta E_I}$ is the partition function.

The phase transition can also be studied by using the probability of the system being in the ferromagnetic state, $P_{\text{FM}}$, as an order parameter, as studied in \cite{Kim2011} using a trapped ion quantum computer. We leave these calculations to a future study.
\subsection{Symmetry of the system}
\label{sec:symmetry}
Next, we discuss the symmetry of the system and explain how it can be utilized to reduce the number of steps in quantum computations. 

A conserved quantity of the system is parity, $(-)^F$, where
\be F = \sum_{i=1}^{N_s} n_i~. \ee
is the total occupation number.
Indeed, it is easy to check that parity commutes with the Hamiltonian \eqref{PBCHam},
\be [ (-)^F , H] = 0~. \ee
Therefore all eigenstates of the Hamiltonian have definite parity, starting with the ground state that has even parity ($(-)^F = +1$).

The Hamiltonian \eqref{PBCHam} is also symmetric under permutations of the sites, $\mathcal{P} : i \mapsto (i+1)\text{mod} N_s$, and reflection around, say, $i=0$, $\mathcal{R} : i \mapsto (-i) \text{mod} N_s$. If $|\psi_I\rangle$ is an eigenstate of the Hamiltonian (eq.\ \eqref{eq:3}), then $\mathcal{P}|\psi_I\rangle$ and $\mathcal{R} |\psi_I\rangle$ are also eigenstates of $H$ belonging to the same eigenvalue $E_I$. If the energy level $E_I$ is non-degenerate, then the corresponding eigenstate must be invariant under permutation and reflection of the sites. Moreover, since $\mathcal{R}^2 = \mathbb{I}$, each energy level consists of states which are either even or odd under reflection of the spatial sites. 

Let us first consider the case $N_s =3$. The ground state must be parity and reflection even. Since the ground state is non-degenerate, it must also be invariant under permutation of the sites. It follows that it has to be of the form
\be |\psi_0\rangle = a|000\rangle + b ( |011\rangle + |101\rangle + |110\rangle )~. \ee
\fix{There is also an excited state of this form but with different coefficients.

Other excited states are obtained by flipping all three qubits in the above expression,
\be  a'|111\rangle + b'( |100\rangle + |010\rangle + |001\rangle )~. \ee
Next, consider an excited state which is odd under parity and reflection. These properties are incompatible with symmetry under permutation of sites, indicating that the energy level is degenerate. It is a double degeneracy with the space spanned by $\{ |\psi_1\rangle , \mathcal{P} |\psi_1\rangle \}$ ($\mathcal{P}^2 |\psi_1\rangle$ is a linear combination of the other two states, since $\mathcal{P}^3 = \mathbb{I}$, and so $\mathcal{P}^2 = - \mathcal{P} - \mathbb{I}$). We may choose
\be |\psi_1\rangle = \frac{1}{\sqrt{2}} (|001\rangle - |010\rangle ) \ee
so that $\mathcal{P} |\psi_1\rangle = \frac{1}{\sqrt{2}} (|100\rangle - |001\rangle )$. Thus, we were able to determine the states of an excited level solely from symmetry considerations.

By the same token, there is another degenerate energy level which is obtained by flipping all three qubits, with states 
\be |\psi_2\rangle = X_0X_1X_2 |\psi_1\rangle = \frac{1}{\sqrt{2}} (|110\rangle - |101\rangle ) \ee
and $\mathcal{P} |\psi_2\rangle = \frac{1}{\sqrt{2}} (|011\rangle - |110\rangle )$.}

For $N_s =4$, the ground state is of the form
\bea  &=& a|0000\rangle + b(|0011\rangle + |0110\rangle + |1001\rangle + |1100\rangle) \nonumber\\ &&+ c  (|0101\rangle + |1010\rangle) + d |1111\rangle~, \eea
easily checked to be parity and reflection even, as well as invariant under permutation. There is an \fix{excited state} of the same form as the ground state and orthogonal to it.

Another excited state is of the form
\bea\label{eq:18}  &=& a(|0001\rangle + |0010\rangle + |0100\rangle + |1000\rangle ) \nonumber\\ &&+ b( |0111\rangle + |1011\rangle + |1101\rangle + |1110\rangle )~, \eea
which is parity odd, reflection even, and invariant under permutation.

There is a degenerate energy level spanned by the states
\be \frac{1}{\sqrt{2}} (|0001\rangle - |0100\rangle) \ , \ \ \frac{1}{\sqrt{2}} (|0010\rangle - |1000\rangle) \label{n2_eig}\ee
which are parity and reflection odd.

Another \fix{excited state} is of the form
\bea && a ( |0001\rangle - |0010\rangle + |0100\rangle - |1000\rangle ) \nonumber\\
&+& b ( |0111\rangle - |1011\rangle + |1101\rangle - |1110\rangle )~, \eea
which is parity odd, reflection even, invariant under permutation and orthogonal to the excited state \eqref{eq:18}.

Another degenerate energy level is spanned by the states
\bea &&\frac{1}{\sqrt{2}} ( |0101\rangle - |1010\rangle ) \ , \ \ \frac{1}{\sqrt{2}} ( |0011\rangle - |0110\rangle ) \ , \nonumber\\ &&\frac{1}{\sqrt{2}} ( |0110\rangle - |1001\rangle ) \ , \ \ \frac{1}{\sqrt{2}} ( |1001\rangle - |1100\rangle )
\label{zero_eig}\eea
all of even parity.

Another set of parity and reflection odd, degenerate higher energy level states are
\be \frac{1}{\sqrt{2}} (|1110\rangle - |1011\rangle) \ , \ \ \frac{1}{\sqrt{2}} (|1101\rangle - |0111\rangle)~. \label{2_eig}\ee
To access the states of the remaining energy levels, it is advantageous to flip the sign and use $-H$ as the Hamiltonian and start by computing its ground state which corresponds to the highest energy level of $H$. The same symmetry considerations apply to the Hamiltonian with flipped sign, $-H$, and one obtains expressions for the higher-level states of $H$ that are similar to the lower-level states obtained above.



\section{Algorithms}
\label{sec:algo}
As mentioned earlier, to calculate the energy levels and corresponding eigenstates of our system we will use a hybrid quantum-classical method based on the QLanczos algorithm which uses the QITE algorithm first proposed in \cite{Motta2019}. Therefore, in this Section we will give a brief overview of these quantum algorithms.

\fix{\subsection{Quantum Imaginary Time Evolution (QITE)}}
\label{sec:QITE}
We start by discussing the QITE algorithm whose classical counterpart was introduced in order to simulate the dynamics of many-body systems. It is advantageous to separate the Hamiltonian into local, but non-commuting, components, $H=\sum_m h_m$. The number of these local terms in the Hamiltonian scales polynomially with the number of particles in the many-body system. Since we are only dealing with a small number of qubits, there is no need to split the Hamiltonian in our case.

QITE relies on evolution in imaginary time. To implement it, we need to set $t\to -i\beta$ in eq.\ \eqref{eq:4} and define the imaginary-time evolution operator $\mathcal{U}=e^{-\beta H}$ which is no longer unitary. 
Starting with the state $|\Psi_0 \rangle$, the evolved state is found in $n$ steps each evolving the system in imaginary time $\Delta\tau$, where $n=\frac{\beta}{\Delta \tau}$,
\be
|\Psi (\beta)\rangle = c_n \left( e^{-\Delta \tau H}  \right)^n |\Psi_0\rangle~,
\ee
with $c_n$ being a normalization constant ($c_n^{-2}=\langle \Psi_0|\mathcal{U}^{2} |\Psi_0\rangle$). In the zero-temperature limit ($\beta\to\infty$), this state converges to the ground state of the system.

The QITE algorithm simulates this non-unitary imaginary-time evolution by approximate unitary updates. Thus, the $s$th step of the imaginary-time evolution,
\be
|\Psi_s\rangle = \frac{c_{s}}{c_{s-1}} e^{-\Delta \tau H} |\Psi_{s-1}\rangle~,\label{sthstep}
\ee
with $s=1, 2, \dots, n$ and $c_0=1$, can be approximated as
\be\label{eq:23}
|\Psi_s\rangle \approx e^{-i\Delta \tau A[s]}|\Psi_{s-1}\rangle~,
\ee
where $A[s]$ can be written in terms of Pauli operators $(\sigma \in \{  X, Y, Z\})$ involving $N_s$ qubits as
\be\label{eq:As}
A[s]=\sum_{i_1, \dots, i_{N_s}} a[s]_{i_1 \dots i_{N_s}} \sigma_{i_1} \dots \sigma_{i_{N_s}}~.
\ee
Once the $a[s]$ coefficients are calculated, these unitary updates can be implemented on a quantum computer. These coefficients can be calculated up to order $\mathcal{O}(\Delta \tau^2)$ by solving a linear system of equations $({\bm{\mathcal{S}} + \bm{\mathcal{S}}^T)\cdot \bm{a}}={\bm{b}}$,
where
\be
\mathcal{S}_{\mathcal{I},\mathcal{I}'} =\langle \sigma_{i_1}\dots \sigma_{i_{N_s}} \sigma_{i'_1}\dots \sigma_{i'_{N_s}} \rangle~,
\ee
and
\be
b_{\mathcal{I}} =-i\sqrt{\frac{c_{s-1}}{c_{s}}} \langle \sigma_{i_1} \dots \sigma_{i_{N_s}} H \rangle~,
\ee
with $\mathcal{I}= \{ i_1,\dots, i_{N_s} \}$, and the expectation values evaluated at the state computed in the previous step, $|\Psi_{s-1}\rangle$. These expectation values involve strings of Pauli matrices and can be evaluated with quantum algorithms recursively. By solving this linear system of equations classically, we obtain the minimum distance between $|\Psi_s\rangle$ and the unitary update \eqref{eq:23} to lowest order in $\Delta\tau$ \cite{Motta2019}. A solution of the linear system of equations can also be found with a quantum algorithm, but we will not do this here as our focus is implementation on NISQ hardware. This kind of quantum algorithm would require implementation of unitary operations with a circuit depth that NISQ hardware could not handle.

In our previous work \cite{Yeter2019}, we found out that these unitary updates for the systems we considered were in the form of a unitary coupled cluster (UCC) \textit{Ansatz}. This is also the case for the current Ising spin chain model. 

The initial state $|\Psi_0\rangle$ determines which eigenstate of the system the QITE algorithm will converge to. It will converge to the ground state as long as $|\Psi_0\rangle$ has a finite overlap with it. For convergence to an excited state, $|\Psi_0\rangle$ must be orthogonal to the ground state.  As we discussed in Section~\ref{sec:symmetry} above, utilizing the symmetry of the system helps us make an educated choice of initial state. In our Ising model, we can exploit the parity and reflection symmetries to choose an initial state for QITE that will be orthogonal to low-level states and therefore converge to the desired energy level. This minimizes the number of required calculations. 

The vector ${\bm{b}}$ has $3^{N_s}$ elements and ${\bm{\mathcal{S}}}$ is a $3^{N_s}\times 3^{N_s}$ matrix, therefore we need to perform $3^{N_s}(3^{N_s}+1)$ measurements in order to calculate all elements in $\bm{b}$ and $\bm{\mathcal{S}}$. Since the Hamiltonian is real, so are these matrix elements. Therefore, in the calculation of ${\bm{b}}$, only the elements that have an odd number of $Y$ Pauli matrices will contribute while the rest will vanish. Similarly, the $\bm{\mathcal{S}} +\bm{\mathcal{S}}^T$ matrix elements which have an even number of $Y$ Pauli matrices will not contribute. Additionally, the $\bm{\mathcal{S}} + \bm{\mathcal{S}}^T$ matrix is symmetric and its diagonal elements are all the same. Using this information, we can reduce the number of measurements significantly.

\fix{As explained in more detail in the next section, although all of the above steps can be performed on quantum hardware, in view of limited resources, we first computed the coefficients $a[s]$ in the unitary updates using quantum simulation. We then implemented the unitary updates with a quantum circuit that produced $|\Psi_s\rangle$ from $|\Psi_0\rangle$ on quantum hardware, aided by the \textit{initialize} function in the IBM Qiskit library. The partial use of quantum simulation limited the error produced by quantum hardware.} If all steps are implemented on NISQ hardware, then the error we are reporting here will be larger and depend on the NISQ device used.

\fix{\subsection{Quantum Lanczos (QLanczos) Algorithm}}
\label{sec:QLanczos}

Next, we apply the QLanczos algorithm which uses the measurement outcomes of the QITE algorithm \fix{in order to obtain} 
\fix{all the} eigenstates of the system \fix{, including excited states}. 
The classical Lanczos algorithm uses the Krylov space $\mathcal{K}$ spanned by a set of vectors $\{|\Phi\rangle, H|\Phi\rangle, H^2|\Phi\rangle, \dots\}$. In its quantum version (QLanczos), $\mathcal{K}$ is spanned by $\{ |\Phi_0\rangle , |\Phi_2\rangle, \dots \}$, where $|\Phi_l \rangle\in \{ |\Psi_s\rangle : s=0,1,\dots\}$.

The number of required QLanczos states $|\Phi_l\rangle$ in the Krylov space is determined by the number of eigenstates of the system that have non-zero overlap with the initial state, $|\Psi_0\rangle$. Our numerical calculations showed that having a smaller number of QLanczos states in the Krylov space than the number of eigenstates with non-zero overlap will result in convergence if we sample from states at a high number of QITE steps (large $s$). 
\begin{figure*}[ht!]
    \centering
    \includegraphics[scale=0.6]{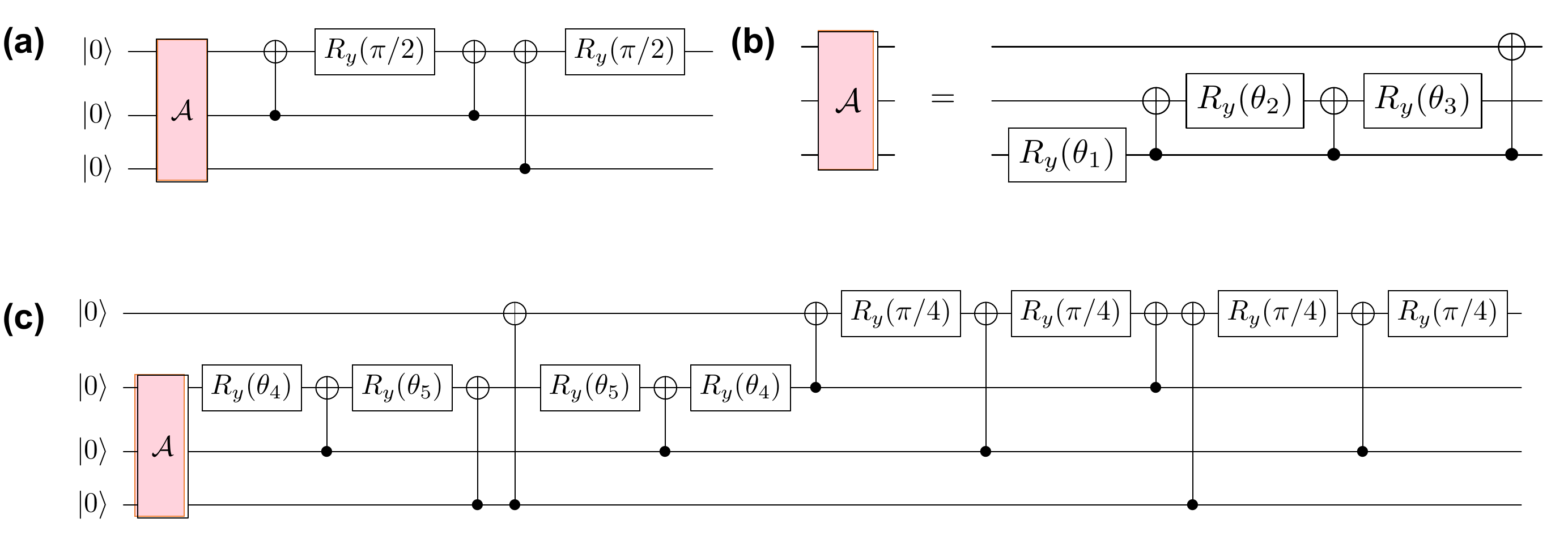}
    \caption{Typical quantum circuits for unitary updates  $|\Psi_s\rangle$ obtained with the aid of the IBM Qiskit \textit{initialize} function. The energy expectation value at each QITE step is obtained from measurements on these quantum circuits. {\bf{(a)}} A 3-qubit quantum circuit. {\bf{(b)}} The 3-qubit gate used in the 3- and 4-qubit quantum circuits expressed in terms of $R_y(\theta)$ rotation and CNOT gates. {\bf{(c)}} A 4-qubit quantum circuit.}
    \label{fig:Qcircuits}
\end{figure*}

After filling the Krylov space with QLanczos states obtained from QITE,
we form the overlap $(\mathcal{T})$ and Hamiltonian $(\mathcal{H})$ matrices whose elements can be calculated in terms of the energy expectation values \fix{obtained from quantum hardware (QITE)}, respectively, as
\be\label{eq:26}
\mathcal{T}_{l,l'}=\langle \Phi_l | \Phi_{l'} \rangle = \frac{c_l c_{l'}}{c_r^2}~,
\ee
and
\be\label{eq:27}
\mathcal{H}_{l,l'}=\langle \Phi_l|H  |\Phi_{l'}\rangle = \mathcal{T}_{l,l'}\langle \Phi_r|H|\Phi_r\rangle~,
\ee
where $r=\frac{l+l'}{2}$, and $l,l'$ are even. The normalization constants can be calculated recursively in terms of expectation values using
\be
\frac{1}{c_{r+1}^2}=\frac{\langle \Phi_r|e^{-2\Delta \tau H}|\Phi_r \rangle}{c_r^2}~. \label{cl}
\ee
\fix{For experimental computation of the normalization constants we expanded $\langle \Phi_r|e^{-2 \Delta \tau H}|\Phi_r\rangle\approx 1-2\Delta \tau \langle \Phi_r|H|\Phi_r\rangle+\mathcal{O}(\Delta \tau^2)$ while keeping in mind that $c_0=1$. The expectation value $\langle \Phi_r|H|\Phi_r\rangle$ was calculated on a quantum computer experimentally using the states generated by the QITE algorithm. The details as to how we obtained the QITE states can be found in Section~\ref{sec:QProgram}. For more accurate results, one can also measure the expectation values for higher powers of the Hamiltonian. The experimentally calculated normalization constants were then used to obtain the matrix elements \eqref{eq:26} and \eqref{eq:27}.}
Thus, all matrix elements of $\mathcal{T}$ and $\mathcal{H}$ were computed with a quantum circuit as expectation values evaluated in the states generated by the QITE algorithm. We then solved the generalized eigenvalue equation
\be
{\bm{\mathcal{H} x}} =E {\bm{\mathcal{T}x}}~, \label{gen_eig}
\ee 
classically and found approximations to the eigenvalues and corresponding eigenstates of the system Hamiltonian, which depended on the choice of initial state $|\Psi_0\rangle$.
For a given eigenvalue $E$, denote the corresponding eigenvector of $\mathcal{H}$ by $\bm{x}^{(E)} = (x_0^{(E)}, x_1^{(E)}, \dots )^T$. We deduce the approximation to an eigenstate of the Hamiltonian \eqref{PBCHam},
\be
|\Psi[E]\rangle= c_E \left( x_0^{(E)} |\Phi_0\rangle + x_1^{(E)} |\Phi_\fix{2}\rangle + \dots \right)~, \label{QLancEq}
\ee
where $c_E^{-1} = \| \sum_{l=0,1,\dots} x_l^{(E)} |\Phi_l\rangle \|$.
Given the state $|\Psi[E]\rangle$, one can recover the approximation to the corresponding energy level using
\be
E = \langle \Psi[E] |H|\Psi[E] \rangle~. \label{QLancEn}
\ee
\fix{This expression for $E$ is redundant, because we have already derived $E$ from Eq.\ \eqref{gen_eig}. However, due to noise the results for the energy levels deduced from \eqref{gen_eig} are numerically unstable. Thus, to obtain $E$, after obtaining the eigenvector $\bm{x}^{(E)}$ from \eqref{gen_eig} classically, we engineered $|\Psi[E]\rangle$ (Eq.\ \eqref{QLancEq}) by building a quantum circuit that we implemented on quantum hardware and calculated the energy expectation value \eqref{QLancEn} experimentally by performing measurements. The quantum circuit was built using \textit{Quantum Programming Studio}~\cite{QPS} and the hardware noise from CNOT gates was reduced by Richardson extrapolation \cite{Li2017} in which the noise is increased purposefully by introducing double CNOT gates corresponding to the each CNOT gate in the quantum circuit and then the extrapolation of the energy expectation value was calculated to obtain the noiseless energy expectation value.


To avoid spurious energy levels $E$, we computed the uncertainty in energy, $\Delta E = ||H|\Psi[E]\rangle-E|\Psi[E]\rangle||$ and discarded eigenvectors $\bm{x}^{(E)}$ with uncertainty exceeding a certain value $\delta$, by demanding $\Delta E \le \delta$. We used $\delta = 0.8$. 

Even though this process improves the numerical stability and accuracy of the experimental results it adds to the total run time of the classical computation.   

Although the noise introduced by quantum hardware increases as the system size grows, making it hard to avoid numerical instabilities, one can improve the numerical stability of the eigenvalues of the generalized eigenvalue equation \eqref{gen_eig} by applying error mitigation techniques such as Richardson extrapolation at each QITE step, or by increasing the order in the series expansion used in the calculation of the normalization constants in \eqref{cl}, or using a different quantum circuit simplification algorithm than Qiskit's \textit{initialize} function resulting in a shorter quantum circuit with fewer CNOT gates. Work in this direction is in progress.
}

\section{Quantum Program}
\label{sec:QProgram}

To calculate the time evolution of various physical quantities, we need the eigenvalues and eigenstates of the system. In our previous work, we demonstrated the practical calculation of the energy spectrum of many-body chemical and nuclear systems by implementing the QITE/QLanczos algorithm on NISQ devices \cite{Yeter2019}. Here, we extend our work to the calculation of energy levels and corresponding eigenstates of the Ising model Hamiltonian \eqref{PBCHam}.

Since the QLanczos algorithm makes use of output from the QITE algorithm, we start with the calculation of energy expectation values of imaginary-time evolution with different initial states informed by symmetry considerations of the system. Using the QITE algorithm outlined above, we calculate the unitary updates (eqs.\ \eqref{eq:23} and \eqref{eq:As}) at every imaginary-time step using a small value of the imaginary-time parameter $\Delta \tau$ and the Hamiltonian \eqref{PBCHam}. Starting with the state $|\Psi_0\rangle$, after $s$ unitary updates, we obtain the state
\be\label{eq:Psis} |\Psi_s\rangle = e^{-i \Delta\tau A[s]} e^{-i \Delta\tau A[s-1]} \cdots e^{-i \Delta\tau A[1]} |\Psi _0\rangle \ee
which we implement with a quantum circuit. We simplified these circuits following the methods discussed in \cite{Shende2006}, as implemented with the \textit{initialize} function in the IBM Q Qiskit library. Examples of 3- and 4-qubit quantum circuits for the states \eqref{eq:Psis} are depicted in Fig.\ \ref{fig:Qcircuits} in terms of single-qubit rotation gates $R_y(\theta)$ and two-qubit CNOT gates. At every imaginary-time step, the angles change, as they depend on the state $|\Psi_s\rangle$, but the depth of the circuit remains the same. Therefore, in terms of economizing the number of gates and operations in the quantum circuit, our results are similar to those in our earlier work \cite{Yeter2019}. It should be noted that, depending on the topology of the quantum hardware, interactions between physical qubits matching those in the quantum circuit implementing \eqref{eq:Psis} may not be readily available, necessitating the addition of SWAP gates to the circuits in Fig.\ \ref{fig:Qcircuits}. 

\fix{\subsection{Error Mitigation}}
\label{sec:err}

Running the quantum circuits on NISQ devices brings errors of various sources such as noise from the implementation of the circuit gates and noise due to the measurement readout errors. To mitigate these errors in the measurements error mitigation strategies are employed. In this work, we only use a readout error mitigation technique \fix{in calculation of the energy expectation values at each QITE step.} 
One can use further error mitigation strategies such as Richardson extrapolation as we did in ref.~\cite{Yeter2019}
or reduced density matrix purification (\cite{McCaskey2019}) to improve the results obtained using the QITE algorithm. 

In this paper, we use local readout error mitigation strategy that we used in our previous work \cite{Yeter2019} in which the corrected expectation values of the Pauli terms is calculated using 

\be
\begin{split}
\label{eq:roem}
\expect{Z_i \dots Z_j}=&\sum_{ x \in {\text{possible outcomes}}}p(x)\\& \ \ \ \ \ \times\frac{(-1)^{x_{i}}-p_i^-}{1-p_i^+}\times\dots \times\frac{(-1)^{x_{j}}-p_j^-}{1-p_j^+}~,
\end{split}
\ee
where $p(x)$ is the probability of each qubit outcome and it takes $2^{N}$ values. Here, we only consider the expectation values for $Z$ terms since we do the measurements in $Z$ basis. The terms with $X$ and $Y$ Pauli operators are rotated to be measured in $Z$ basis. We define the symmetric and anti-symmetric combinations of the probability of $i$-th qubit flipping from 0 to 1 ($p_i(0|1)$) or from 1 to 0 ($p_i(1|0)$) as
\be
p_i^\pm=p_i(0|1)\pm p_i(1|0)~,
\ee
with
\be
p(1|0)= \frac{\# \ \text{of states expected in $|1\rangle$ measured in $|0\rangle$}}{\# \ \text{of shots}}
\ee
or vice versa for $p(0|1)$. 



\begin{figure}[ht!]
    \centering
    \includegraphics[scale=0.40]{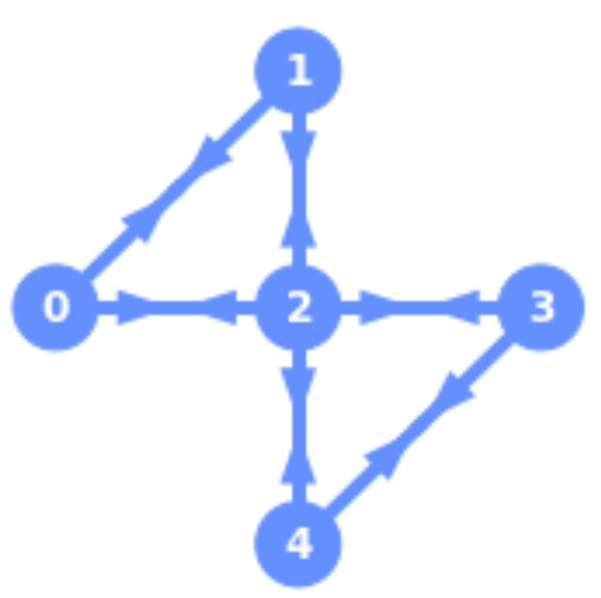}
    \caption{The quantum circuits \fix{for QITE algorithm} were run on 5-qubit IBM Q Yorktown (version v2.0.5) hardware because of its periodic topology. The arrows in the figure indicate the direction of the CNOT gates.}
    \label{fig:Yorktownlayout}
\end{figure}
\
\
\section{Results and Discussion}
\label{sec:Results}

\fix{\subsection{QITE and QLanczos Results}}
The experiments \fix{for the QITE algorithm} were run on 5-qubit IBM Q Yorktown hardware. The number of shots for the each experiment was 8192 and each experiment was run $N_{\text{runs}}=3$ times to calculate the statistical error in the measurements. The reason for choosing this quantum computer out of other IBM Q's cloud accessible devices is its periodic topology as seen in Fig. \ref{fig:Yorktownlayout}. Using a quantum computer with periodic topology reduces the number of required SWAP gates for our periodic Ising spin chain Hamiltonian which reduces the number of required CNOT gates. This is important because CNOT gates are the dominant source of the error in a quantum circuit. For comparison, Honeywell's ion trap quantum computer offers connectivity between all physical qubits. Therefore, the error in this type of quantum system might be smaller since it does not require the addition of SWAP gates for the type of interaction Hamiltonian considered here. The basis gates which can be directly implemented on IBM Q Yorktown quantum computer are single-qubit gates $U$ and the two-qubit CNOT gate, where
\be
U(\theta,\phi,\lambda)=\begin{pmatrix}
    \cos{\frac{\theta}{2}} & -e^{i\lambda}\sin{\frac{\theta}{2}} \\
    e^{i\phi}\sin{\frac{\theta}{2}} & e^{i(\phi+\lambda)}\cos{\frac{\theta}{2}}
\end{pmatrix}~,
\ee
is a general three-parameter single-qubit gate.
In Fig.\ \ref{fig:Qcircuits}, we used the single-qubit rotation gate $R_y(\theta)$ which can be expressed in terms of the basis gates as $R_y(\theta)=U(\theta,0,0)$.

\begin{figure}[ht!]
    \centering
    \includegraphics[scale=0.60]{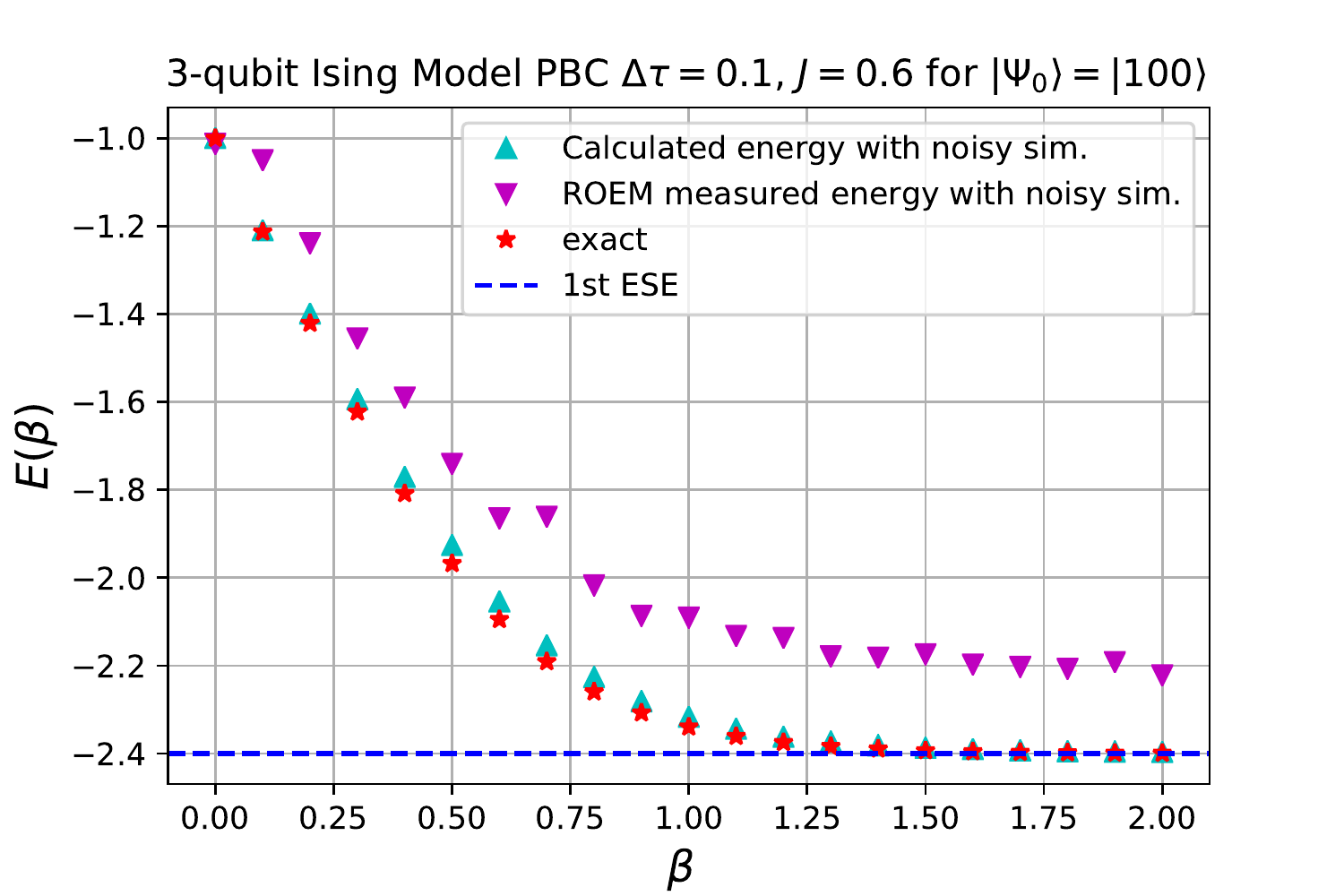}
    \caption{Energy \textit{vs.}~imaginary time calculated exactly and compared to the one calculated using a noisy simulator, and ROEM measured energy from the noisy hardware of IBM Q Yorktown. Initial state is $|100\rangle$. The parameters are set to $h_T=1$ and $J=0.6$. Imaginary-time step is $\Delta\tau = 0.1$. Energies converge to energy level $-2.4$.}
    \label{fig:QITEmeasurement}
\end{figure}

As mentioned in the \fix{Section~\ref{sec:QITE}}, simulating each QITE step requires significant number of measurements on hardware. Even using the aforementioned properties of $\bm{b}$ and ${\bf{\mathcal{S}}}$ matrices there needs to be a large number of measurement done to apply the QITE algorithm on hardware. For example, for $N_s=3$ we were able to reduce the number of measurements from 756 to 187 at every QITE step.  Due to limitations in cloud access to the quantum hardware (such as long queue and connection interruptions) we simulated the quantum circuits for the states $|\Psi_s\rangle$ and implemented them on quantum hardware to obtain the energy expectation values for various values of imaginary time. With full implementation on a NISQ device, additional errors will occur. To estimate these additional errors, we considered a generic case and fully implemented it on simulated quantum hardware. We obtained energy expectation values for various values of imaginary time for three sites, $N_s=3$, using the initial state $|\Psi_0\rangle=|100\rangle$, and the Ising model with parameters $J=0.6$ and $h_T=1$. We implemented the QITE algorithm and obtained the operator $A[s]$ from measurements on the noisy simulator of the same backend. We used $N_{\text{shots}}=8192$ and the calibration parameters from 04/24/2020. In Fig.\ \ref{fig:QITEmeasurement} we compare the convergence of the energy expectation values to the first excited state energy in three different cases, \textit{(a)} from exact calculation of the state $|\Psi_s\rangle$ as well as energy expectation values, \textit{(b)} from a noisy simulation of the state $|\Psi_s\rangle$ and exact energy expectation values, and \textit{(c)} from a noisy simulation of both the state $|\Psi_s\rangle$ and readout error mitigated (notated as ROEM in the rest of the paper) energy expectation values. The energy expectation values obtained using methods \textit{(a)} and \textit{(b)} are very close to each other, showing that the main source of additional error is due to measurements. It follows that the use of simulated states does not introduce significant errors. 
However, the ROEM energy expectation values obtained from measurements on quantum hardware differ from results from noiseless simulations. In what follows, we use \fix{noiseless} simulated states, implement their quantum circuits on quantum hardware, and perform measurements to obtain energy expectation values. 

\begin{figure*}[ht!]
    \centering
    \includegraphics[scale=0.65]{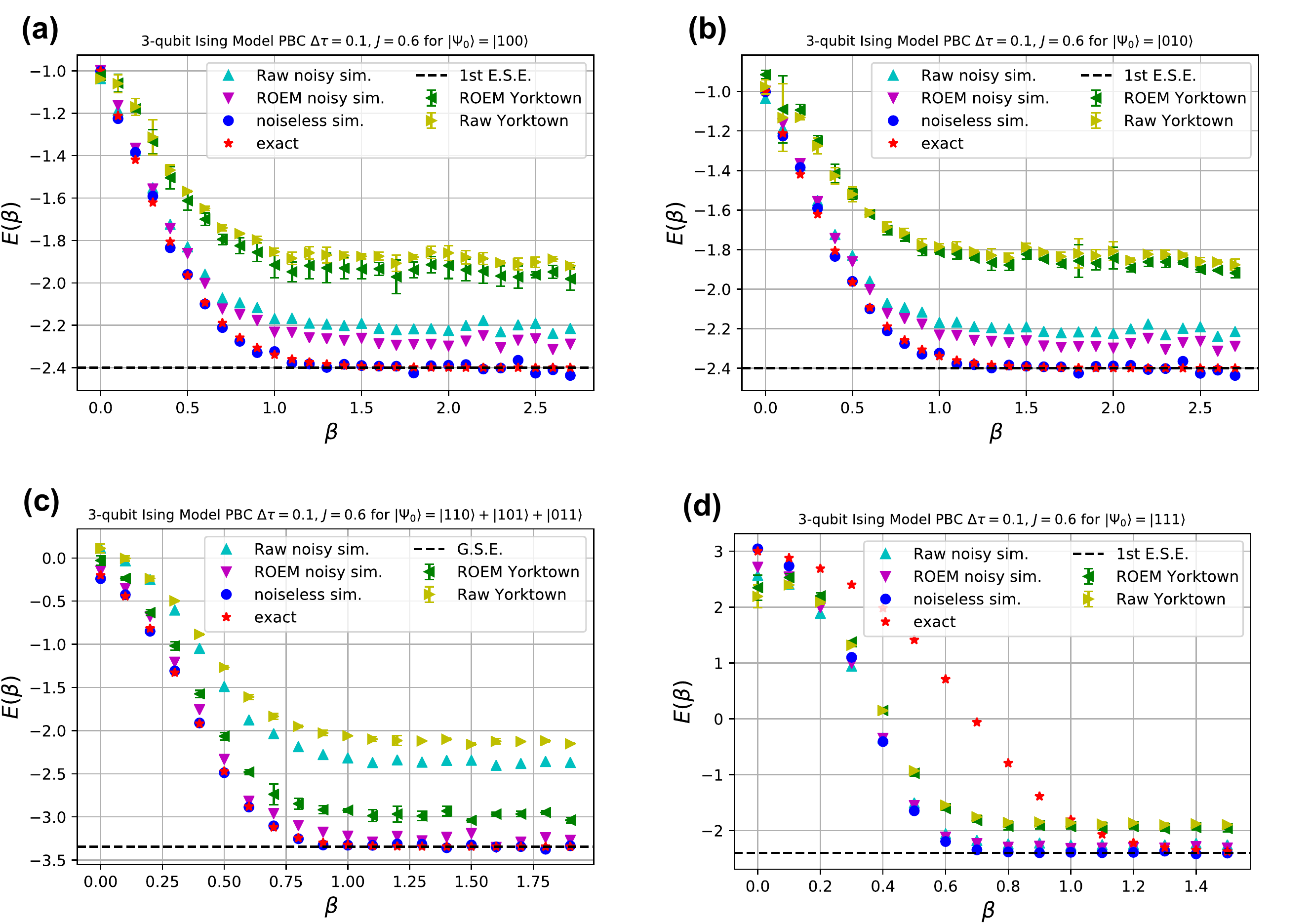}
    \caption{Energy \textit{vs.}~imaginary time calculated exactly using $H$, and compared to IBM Q Aer QASM noiseless and noisy simulator, IBM Q Yorktown hardware raw and ROEM energy expectation values. The initial state is {\bf{(a)}} $|100\rangle$, {\bf{(b)}} $|010\rangle$, {\bf{(c)}} $\frac{1}{\sqrt{3}} (|110\rangle+|101\rangle+|011\rangle)$, {\bf{(d)}} $|111\rangle$. Data on {\bf{(a)}}, {\bf{(b)}} and {\bf{(c)}}, and {\bf{(d)}} collected on 04/23-24/2020, 04/19-22/2020, 05/01/2020 and 04/22-24-25/2020, respectively. For the hardware data, $N_{\text{run}}=3$ and the error bars are $\pm \sigma$. Runs for {\bf{(a)}}, {\bf{(b)}}, {\bf{(d)}} were on qubits $[q_0, q_1, q_2]=[0, 1, 2]$, whereas for {\bf{(c)}} on $[2, 3, 4]$, because on the respective days of the runs, the backend properties were better for those qubits. The parameters are set to $h_T=1$ and $J=0.6$. Imaginary-time step is $\Delta\tau = 0.1$. Energies converge to first-excited energy level $-2.4$ ({\bf{(a)}}, {\bf{(b)}} and {\bf{(d)}}) and ground-state energy level $-3.4$ ({\bf{(c)}}).  }
    \label{fig:3qubit_QITE}
\end{figure*}

\begin{figure*}[ht!]
    \centering
    \includegraphics[scale=0.55]{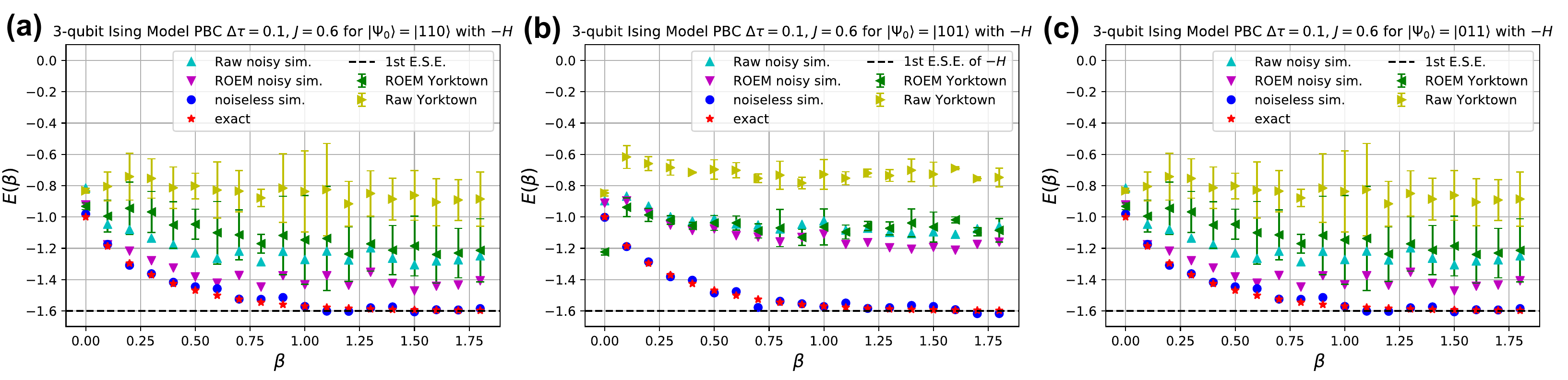}
    \caption{Energy \textit{vs.}~imaginary time calculated exactly using $-H$, and compared to IBM Q Aer QASM noiseless and noisy simulator, IBM Q Yorktown hardware raw and ROEM energy expectation values. The initial state is {\bf{(a)}} $|110\rangle$, {\bf{(b)}} $|011\rangle$, {\bf{(c)}} $|101\rangle$. Data were collected on days 06/12/2020-06/13/2020. For the hardware data, $N_{\text{run}}=3$ and the error bars are $\pm \sigma$. Runs to obtain these data were done on qubits $[q_0, q_1, q_2]=[2, 3, 4]$. The parameters are set to $h_T=1$ and $J=0.6$. Imaginary-time step is $\Delta\tau = 0.1$. Energies converge to energy level $-1.6$.}
    \label{fig:3qubit_QITE_nH}
\end{figure*}

\begin{figure*}[ht!]
    \centering
    \includegraphics[scale=0.60]{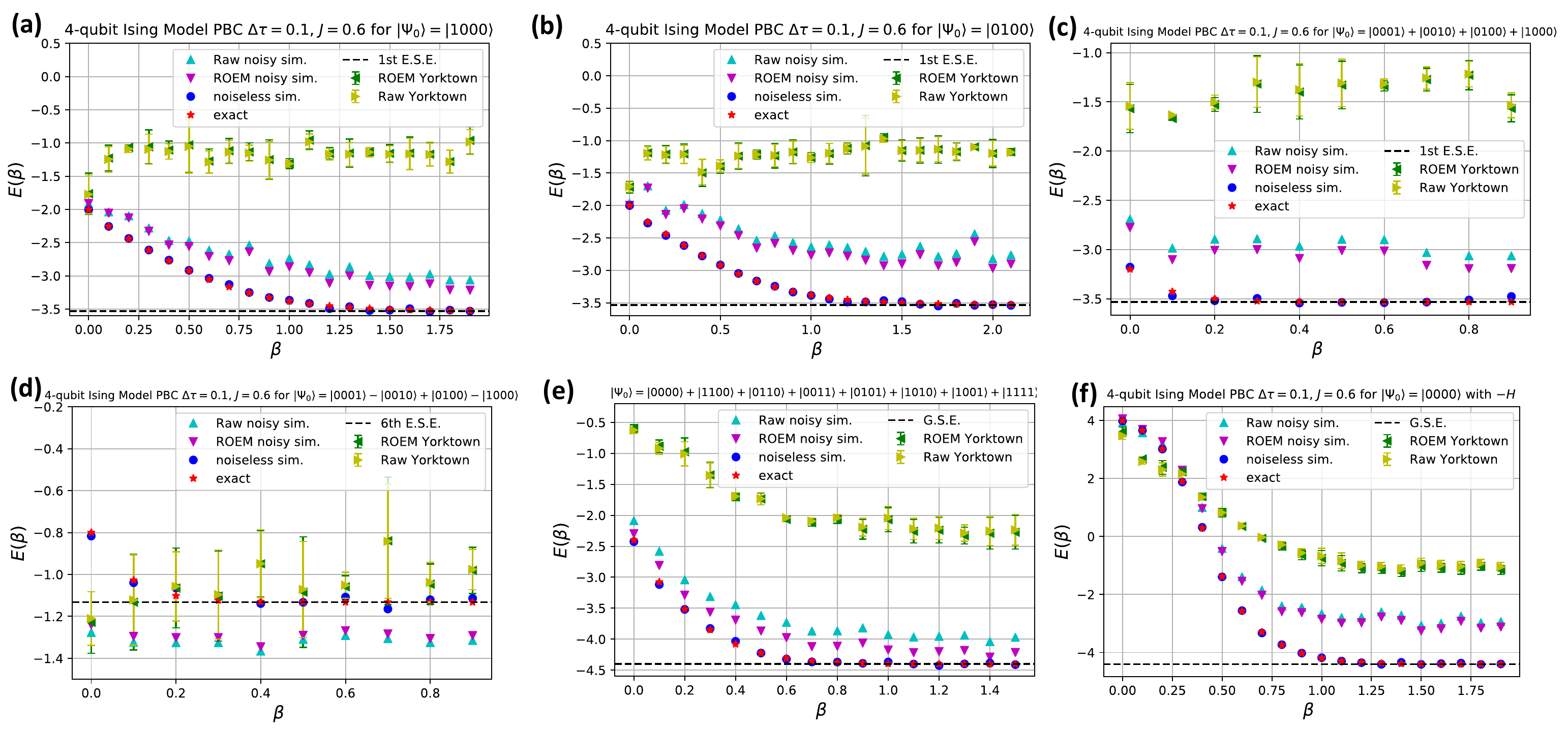} \caption{Energy \textit{vs.}~imaginary time calculated exactly and compared to IBM Q Aer QASM noiseless and noisy simulator, IBM Q Yorktown hardware raw and ROEM energy expectation values. The initial state is {\bf{(a)}} $|1000\rangle$, {\bf{(b)}} $|0100\rangle$, {\bf{(c)}} $\frac{1}{2} (|0001\rangle+|0010\rangle+|0100\rangle+|1000\rangle)$, {\bf{(d)}} $\frac{1}{2} (|0001\rangle-|0010\rangle+|0100\rangle-|1000\rangle)$, {\bf{(e)}} $\frac{1}{\sqrt{7}} (|0000\rangle+|1100\rangle+|0110\rangle+|0101\rangle+|1010\rangle+|1001\rangle+|1111\rangle)$, {\bf{(f)}} $|0000\rangle$ (with $-H$).
    Runs to obtain these data were done on days {\bf{(a)}} 04/26/2020-05/13/2020, {\bf{(b)}} 05/15/2020, {\bf{(c)}} 04/22, 28/2020, {\bf{(d)}} 04/22/2020, {\bf{(e)}} 05/06, 08/2020, {\bf{(f)}} 06/01/2020, respectively using qubits $[q_0, q_1, q_2, q_3]=[0, 1, 2, 3]$. For the hardware data, $N_{\text{run}}=3$ and the error bars are $\pm \sigma$. The parameters are set to $h_T=1$ and $J=0.6$. Imaginary-time step is $\Delta\tau = 0.1$. Energies converge to first-excited energy level $-3.4$ ({\bf{(a)}}, {\bf{(b)}} and {\bf{(c)}}), the ground-state energy level $-4.4$ ({\bf{(e)}} and {\bf (f)}), and the energy level $-1.1$ (\textbf{(d)}).}
    \label{fig:4qubit_QITE}
\end{figure*}

The results of measurements on these quantum circuits produced by the QITE algorithm as a function of imaginary time for different initial states are depicted in Figs.\ \ref{fig:3qubit_QITE}, \ref{fig:3qubit_QITE_nH}, and \ref{fig:4qubit_QITE} for $N_s=3$ and $N_s=4$ spatial sites of our Ising model with parameters $h_{\text{T}}=1$ and $J=0.6$. In the $N_s=3$ case, the initial states $|\Psi_0\rangle$ are chosen as $|100\rangle$, $|010\rangle$, $\frac{1}{\sqrt{3}} (|110\rangle+|101\rangle+|011\rangle)$, and $|111\rangle$ shown in Fig.\ \ref{fig:3qubit_QITE} \textbf{(a)-(d)}, respectively.

Similarly, in the $N_s=4$ case, the initial states are chosen as $|1000\rangle$, $|0100\rangle$, $\frac{1}{2} (|0001\rangle+|0010\rangle+|0100\rangle+|1000\rangle)$, $\frac{1}{2} (|0001\rangle-|0010\rangle+|0100\rangle-|1000\rangle)$, and $\frac{1}{\sqrt{7}} (|0000\rangle+|1100\rangle+|0110\rangle+|0101\rangle+|1010\rangle+|1001\rangle+|1111\rangle)$. They are shown in Fig.\ \ref{fig:4qubit_QITE} \textbf{(a)-(d)}, respectively. These initial states are chosen by taking the symmetry of the system into consideration, as explained in Section~\ref{sec:symmetry}.

The QITE algorithm converges to the minimum of the symmetry group that the initial state belongs to. Therefore, it might be challenging to access higher-value energy levels using the QITE and QLanczos algorithms. To facilitate the algorithm's convergence to higher levels, we reversed the sign of the Hamiltonian \eqref{PBCHam} so that high energy levels turn into low levels whereas the corresponding eigenstates remain the same. We applied this strategy to calculate some of the high energy levels and corresponding eigenstates of our system, e.g., for the 4th and 5th excited states in the 3-qubit ($N_s=3$), and the 15th excited state in the 4-qubit ($N_s=4$) case. The results of this strategy for the QITE algorithm, including energy expectation values, can be seen in Figs.\ \ref{fig:4qubit_QITE}\textbf{(e)} and \ref{fig:3qubit_QITE_nH}. In these examples, since we are looking for the minimum of the reverse Hamiltonian $-H$, we chose the initial states $|\Psi_0\rangle$ to be reflection and parity symmetric, namely $|110\rangle$, $|011\rangle$, $|101\rangle$, and $|0000\rangle$, respectively. 

\begin{figure*}[ht!]
    \centering
    \includegraphics[scale=0.50]{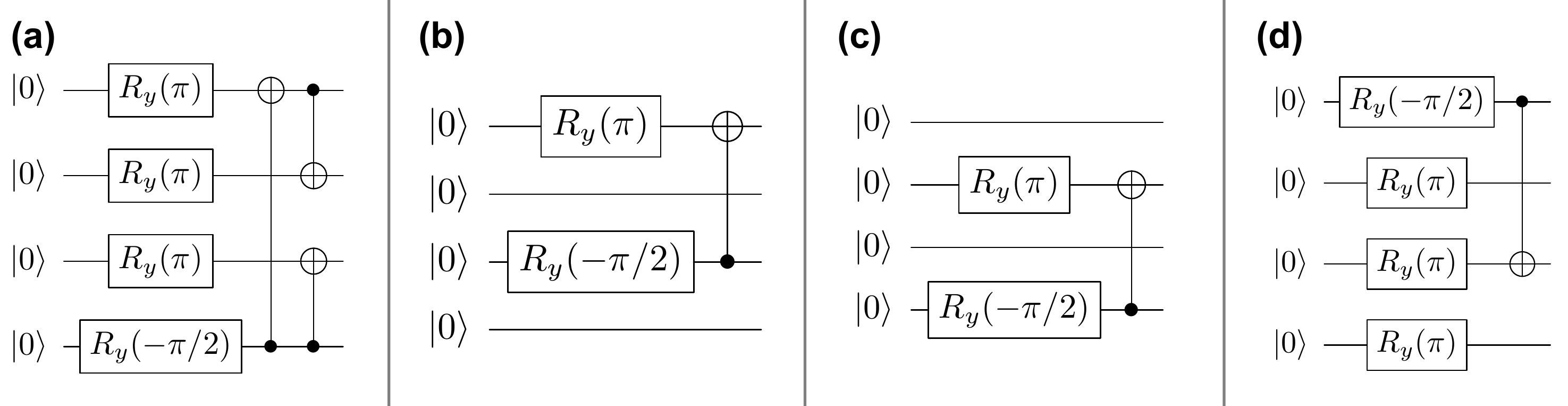}
    \caption{The quantum circuits used to calculate energy levels with exact initial states $|\Psi_0\rangle$: {\bf{(a)}} 0, with $\frac{1}{\sqrt{2}} (|0101\rangle-|1010\rangle)$, {\bf{(b)}} -2, with $ \frac{1}{\sqrt{2}} (|0010\rangle-|1000\rangle)$, {\bf{(c)}} -2, with $ \frac{1}{\sqrt{2}} (|0001\rangle-|0100\rangle)$, and {\bf{(d)}} 2, with $ \frac{1}{\sqrt{2}} (|1110\rangle - |1011\rangle)$. The parameters were set to $J=0.6$ and $h_T=1$.
    }
    \label{fig:qcfor0andn2}
\end{figure*}

As mentioned in Section~\ref{sec:symmetry}, some of the eigenstates are completely constrained by the symmetry of the system, therefore calculating them is redundant. For example, in the $N_s=4$ case for parameters $J=0.6$ and $h_T=1$ the zero eigenvalue is degenerate and the corresponding exact eigenstates are given by eq.\ \eqref{zero_eig}. Similarly, the eigenstates correponding to the degenerate energy level $-2$ are given analytically by eq.\ \eqref{n2_eig}. We took advantage of the exact expressions for these eigenstates in our calculations. Although, we used the exact eigenstates obtained using symmetry constraints, we measured the energy expectation values for the eigenstates demonstrated as the first state in \eqref{zero_eig} and states in \eqref{n2_eig} on hardware (IBM Q Yorktown) using the quantum circuits seen in Fig. \ref{fig:qcfor0andn2} {\bf{(a)}}, {\bf{(b)}}, and {\bf{(c)}}, respectively. These circuits were run on hardware $N_{\text{runs}}=3$ times with each run having $N_{\text{shots}}=8192$ on 08/12/2020 with qubit layout $[q_0, q_1, q_2, q_3 ]=[1,0,3,2]$ and the ROEM average energy values obtained are $0.037 \pm 0.006$, $-2.06 \pm 0.02$, and $-2.01 \pm 0.01$ (where the $\pm$ error is the standard deviation of the mean) compared to the exact eigenvalues of 0 and -2, respectively. For the same coupling and magnetization parameters the states expressed in \eqref{2_eig} correspond to eigenvalue 2 and it is degenerate. Although these states correspond to 3 occupied sites and since we study single particle propagation and two-particle scattering only we did not need them in our calculations we obtained an experimental ROEM mean value of $2.05\pm 0.03$ for the first state in \eqref{2_eig} using the circuit in Fig.~\ref{fig:qcfor0andn2}{\bf{(d)}}. The experiments were run on IBM Q Yorktown, $N_{\text{runs}}=3$ times with each run having $N_{\text{shots}}=8192$ on 08/13/2020 with qubit layout $[q_0, q_1, q_2, q_3 ]=[0,1,2,3]$.


In our current study, we used two-dimensional Krylov spaces. Although, depending on the choice of initial state, convergence might take longer for a low-dimensional Krylov space, adding more dimensions causes numerical instabilities and does not guarantee convergence to eigenstates of the Hamiltonian \eqref{PBCHam}. Interestingly, we were able to observe convergence to the eigenvalues of the system by using a three-dimensional Krylov space together with our uncertainty criterion to exclude spurious states ($\Delta E \le \delta$). However, we did not obtain three distinct energy eigenstates. 
In general, results were numerically more accurate in two-dimensional Krylov spaces for the QLanczos algorithm. Adding more dimensions decreased the number of cases where off-diagonal $\mathcal{T}$ matrix elements were $<1$ resulting in spurious eigenstates. Our numerical results indicate that using two-dimensional Krylov space is the \fix{optimal} choice for \fix{the implementation of the QLanczos algorithm on noisy quantum devices of this particular system with the parameters used in this study.}
Further application of the error mitigation strategies, such as Richardson extrapolation (an example of application of Richardson extrapolation to QITE algorithm can be seen in \cite{Yeter2019}), might improve the numerical stability and can provide faster convergence in higher-dimensional Krylov spaces.

\begin{figure*}[ht!]
    \centering
    \includegraphics[scale=0.5]{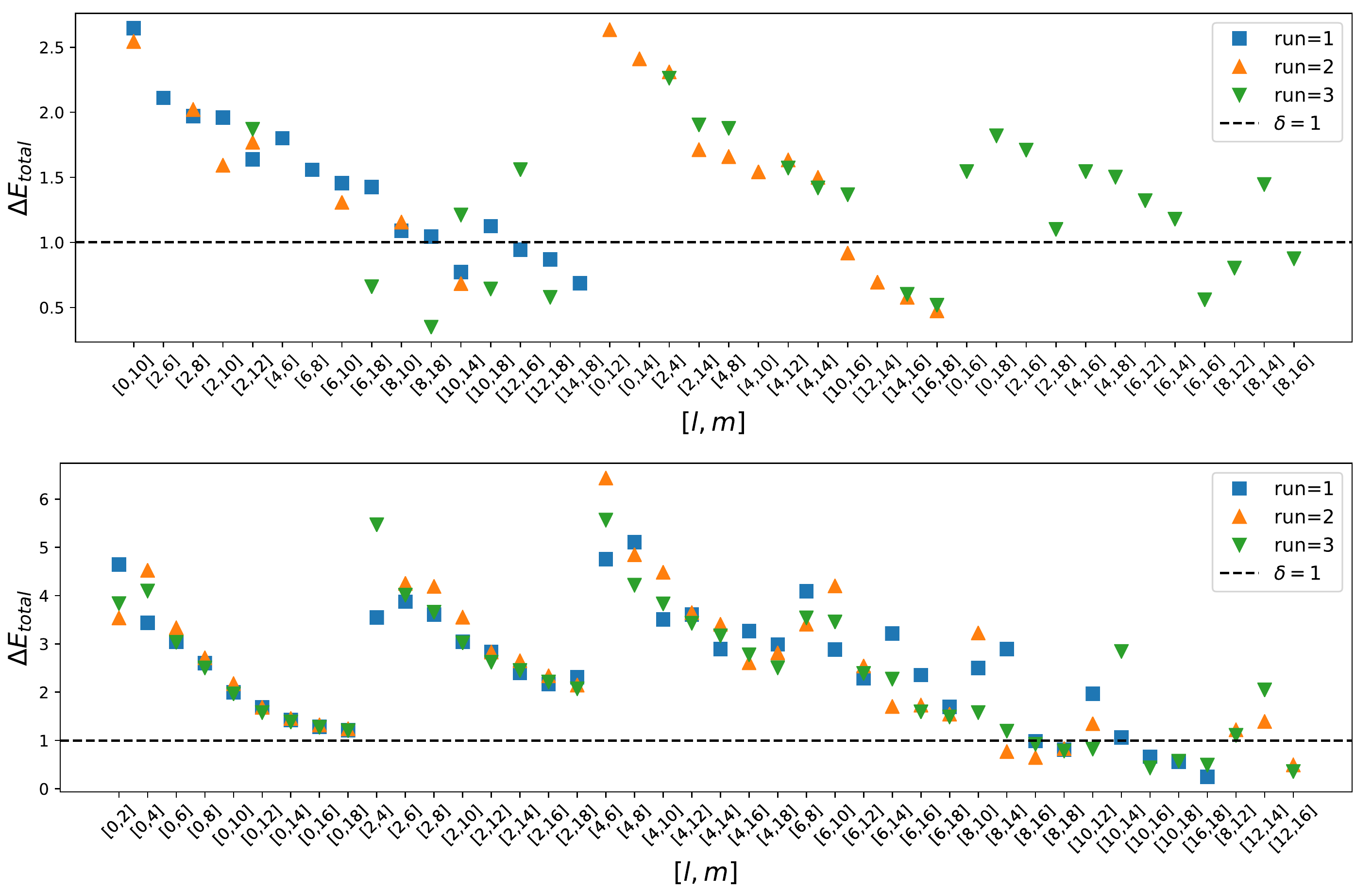}
   \caption{ The uncertainty in energy $\Delta E$ for different Krylov ($\mathcal{K}$) space parameters $[l, m]$, where $\{|\Phi_l\rangle, |\Phi_m\rangle \}$ span $\mathcal{K}$, and initial states {\bf{(a)}} $|1000 \rangle$ (\textit{cf.}~with Fig.\ \ref{fig:4qubit_QITE}{\bf{(a)}}) and {\bf{(b)}} $|0000 \rangle$ (\textit{cf.}~with Fig.\ \ref{fig:4qubit_QITE}{\bf{(f)}}). Experimental results from $N_{\text{runs}}=3$ runs on IBM Q Yorktown hardware. }
    \label{fig:4qubit_norm}
\end{figure*}

As mentioned in Section~\fix{\ref{sec:QLanczos}}, we decide on the convergence to the eigenstates and eigenvalues of the system and discard spurious states by using the uncertainty criterion, $\Delta E \le \delta$. Two examples involving the ground and excited states that are specific to a given initial state, $|\Psi_0\rangle$, are shown in Fig.~\ref{fig:4qubit_norm}. Specifically, the uncertainty $\Delta E$ is shown for $N_s=4$ and various values of $(l,m)$, where $l,m$ are even integers and label the basis states of the Krylov space, which is spanned by $\{|\Phi_l\rangle,|\Phi_m\rangle\}$. Results of the 3 different runs on IBM Q Yorktown hardware are shown. We keep increasing $l$ and $m$, which correspond to QLanczos states with higher QITE steps, 
until $\Delta E < 1$, and we choose the eigenvalues and eigenstates that give the minimum uncertainty. 

\begin{figure*}[ht!]
    \centering
    \includegraphics[scale=0.50]{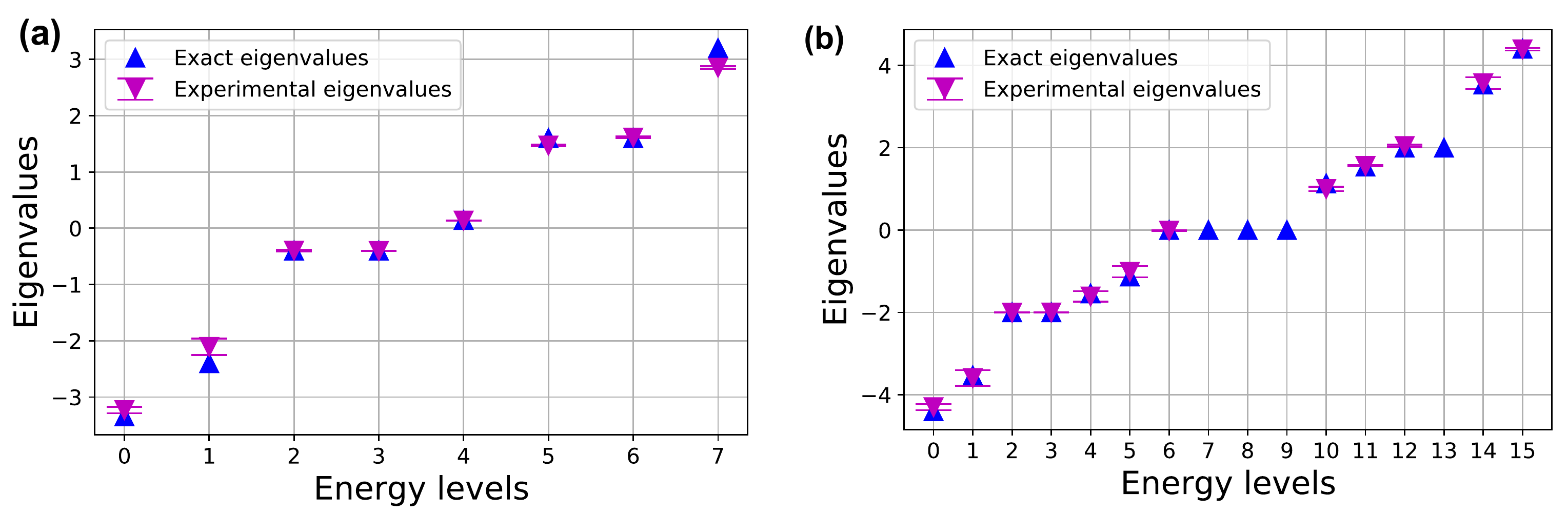}
    \caption{Exact and experimental eigenvalues of the Hamiltonian for $N_s=3$ ({\bf{(a)}}) and $N_s=4$ ({\bf{(b)}}). The parameters are set to $J=0.6$ and $h_T=1$. For $N_s=4$, the 2nd, 3rd, and 6th energy levels were obtained from the exact eigenstates (Eqs.\ \eqref{n2_eig} and \eqref{zero_eig}). The remaining eigenvalues were obtained \fix{from the eigenstate quantum circuits ran on IBM Q Yorktown, Vigo, Casablanca and Manhattan devices.  The Richardson extrapolation error mitigation strategy was applied to reduce effect of the quantum hardware noise.}
    The experiments were run $N_{\text{runs}}=3$ times and the error bars represent one standard deviation. }
    \label{fig:3_4_qubit_eigenvals}
\end{figure*}

\fix{After the application of this process, as explained earlier, we ran each quantum circuit corresponding to the each eigenvector obtained from our hybrid quantum-classical QLanczos algorithm on quantum hardware, specifically on IBM Q Vigo, Casablanca, Manhattan devices depending on their availabilities. This gave us the experimental energy eigenvalues for $N_s=3$ and $N_s=4$ with parameters $J=0.6$ and $h_T=1$ PBC Ising model.   
As a result, using either the exact states obtained from symmetry or using our QLanczos algorithm with a Krylov space of size 2 we obtained experimental energy eigenvalues as seen in Fig.~\ref{fig:3_4_qubit_eigenvals} {\bf{(a)}} ($N_s=3$) and {\bf{(b)}} ($N_s=4$) which are in very good agreement with the exact eigenvalues of the Hamiltonian in \eqref{PBCHam}.}

\fix{\subsection{Time Evolution Results}}
\label{sec:res}
\fix{Finally, we obtained the coefficients $t_{Ix}$ in \eqref{eq:5} from the measurements of each component of the eigenvector. These measurements give the absolute value of each coefficient, $|t_{Ix}|$. Since they are all real, in order to determine them, we need to find the sign. This requires additional measurements with an ancilla qubit, but they introduce no errors because of the binary nature of the sign. We used these coefficients in Eqs.\ \eqref{eq:trans}, \eqref{eq:ni}, and \eqref{eq:mz} to calculate the transition amplitudes, occupation number, and average magnetization as functions of time. 


We summarized our method to calculate the transition amplitudes, occupation number, and transverse magnetization using QLanczos algorithm in the pseudocode below in Fig. \ref{fig:algorithm}. \fix{If the algorithm fails to find eigenvalues and corresponding eigenvectors of the Hamiltonian, then the initial parameters should be changed. If the choice of initial state is informed by symmetry considerations and the uncertainty keeps decreasing at each step, one can start the algorithm with a larger $s_{max}$. In Fig.~\ref{fig:algorithm} we denoted the measurements needed to be performed on quantum hardware as [Q] and the classical computations were denoted by [C]. As discussed earlier, due to constraints in quantum hardware access, we calculated $A[s]$ in Step 3 classically and used the Qiskit $\textit{initialize}$ function to find the quantum circuit in Step 4 of the pseudocode in order to reduce the depth of the circuit.}


\begin{figure}[ht!]
    \centering
    \includegraphics[scale=0.73]{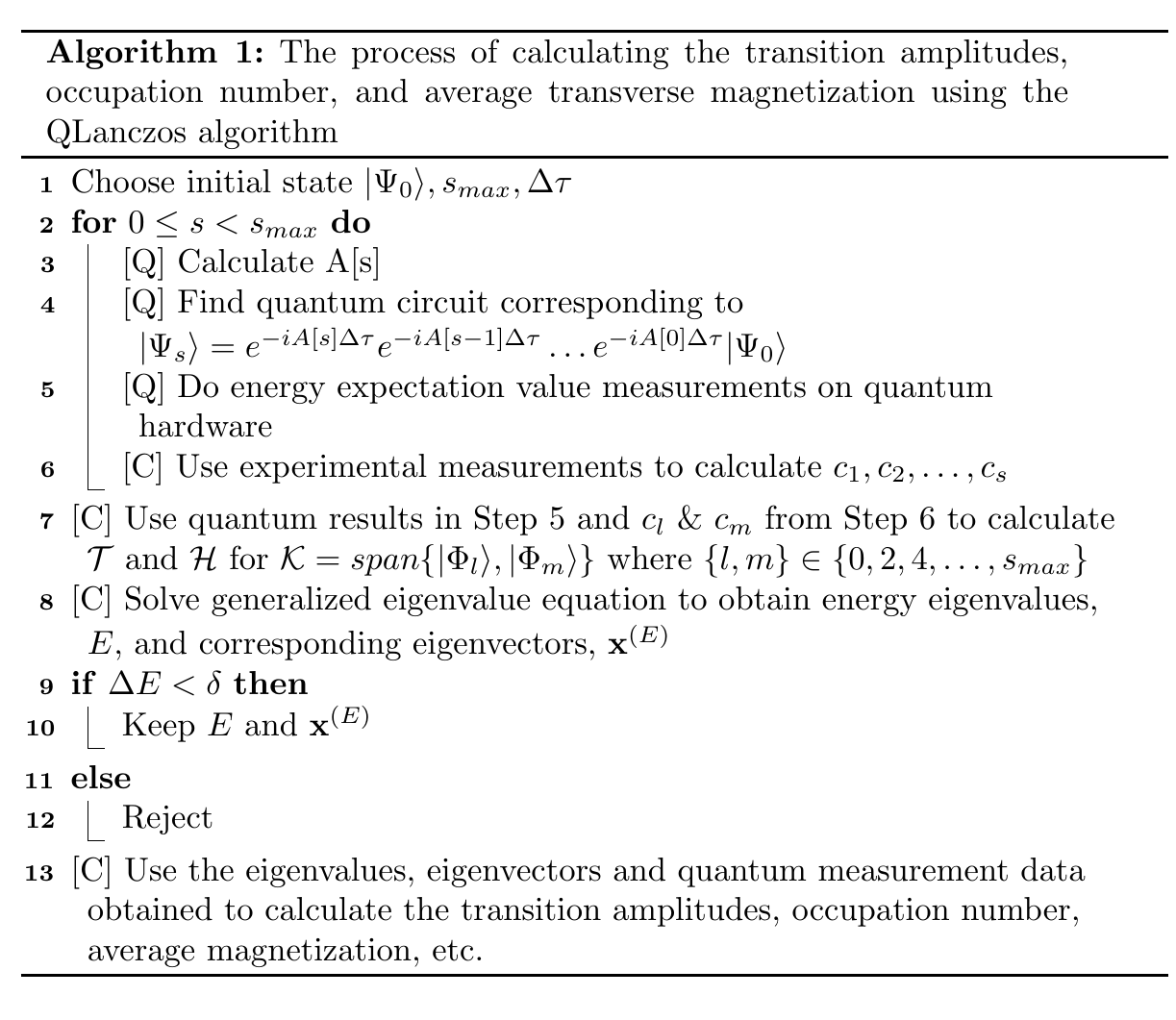}
  \caption{The algorithm where the process of calculating the transition amplitudes, occupation number and average transverse magnetization is summarized. \fix{Here, [Q] indicates the measurements on quantum hardware and [C] indicates classical computation. } }
    \label{fig:algorithm}
\end{figure}}

Here, we present our
experimental data
obtained from data on the IBM Q Yorktown, \fix{Vigo, Casablanca and Manhattan} hardware for transition probability amplitudes
, occupation number at each spatial site,
and average transverse magnetization 
for number of spatial sites $N_s=3$ and $N_s=4$. 

We chose the parameters of the system Hamiltonian in \eqref{PBCHam} to be $h_{\text{T}}=1$ and $J=0.6$.
In ref.\ \cite{Gustafson2019_1} it was found that errors arising from quantum hardware become worse as the coupling $J$ increases. In this section we present results which show small hardware errors even as one moves away from the weak coupling regime.

\begin{figure*}[ht!]
    \centering
    \includegraphics[scale=0.65]{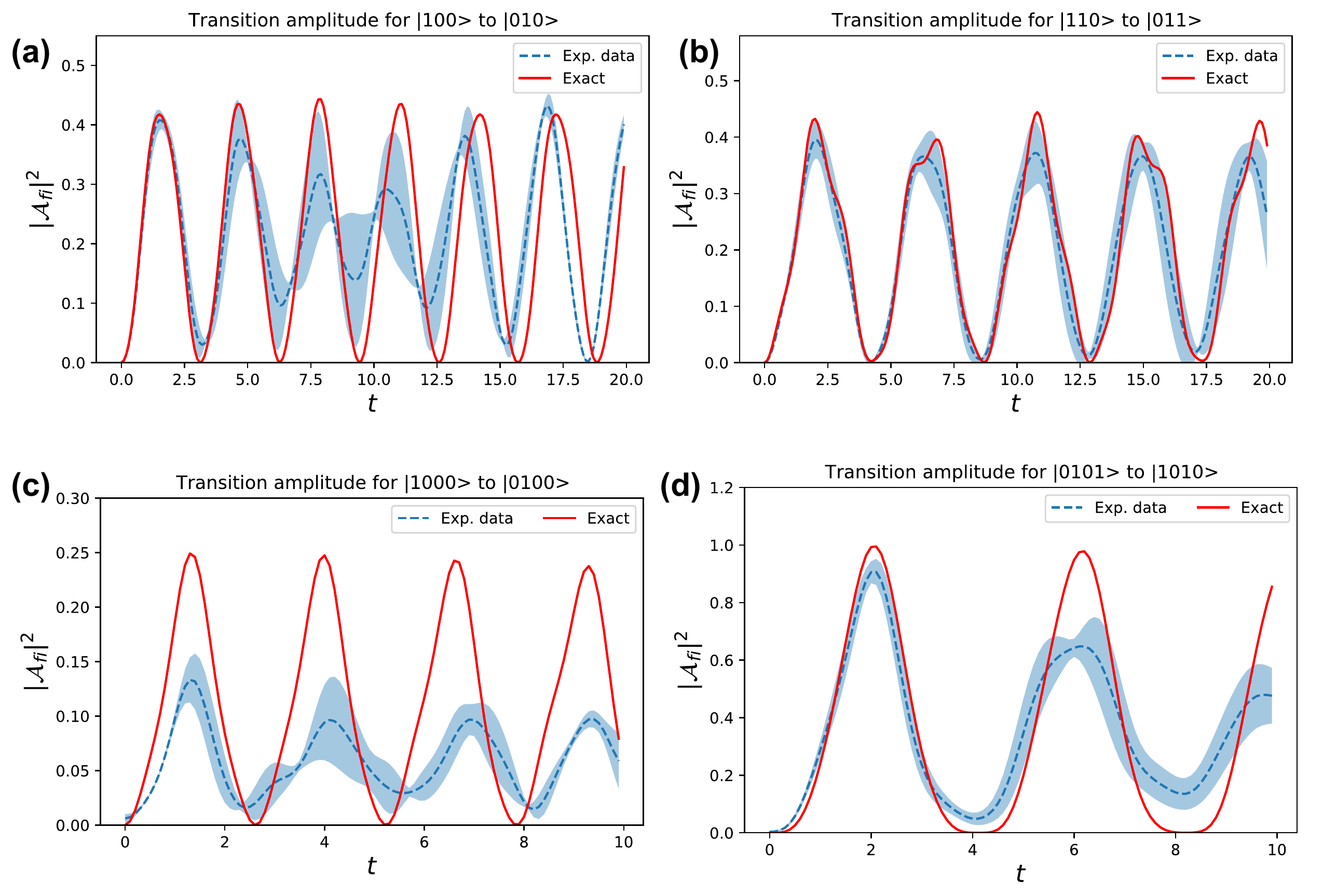}
    \caption{Transition probabilities \textit{vs.}~time calculated using the energies obtained from exact diagonalization and compared to those from ROEM energies using the QLanczos algorithm on IBM Q Yorktown hardware. The transitions are {\bf{(a)}} $|100\rangle \to |010\rangle$, {\bf{(b)}} $|110\rangle \to |011\rangle$, {\bf{(c)}} $|1000\rangle \to |0100\rangle$, and {\bf{(d)}} $|0101\rangle \to |1010\rangle$. The parameters are set to $J=0.6$ and $h_T=1$. $N_{\text{runs}}=3$, and the shaded regions are showing one-standard-deviation error.}
    \label{fig:Transition_average}
\end{figure*}

In Figs.\ \ref{fig:Transition_average} {\bf{(a)}}-{\bf{(d)}}, we show numerical values of the transition amplitudes calculated from given exact $|\text{initial}\rangle$ and $|\text{final}\rangle$ states, and compare them with values obtained from experimental data produced by the QLanczos quantum algorithm that calculates energy eigenvalues and corresponding eigenstates. Figs.\ \ref{fig:Transition_average} {\bf{(a)}} and {\bf{(c)}} show the one-particle propagation probability, and Figs.\ \ref{fig:Transition_average} {\bf{(b)}} and {\bf{(d)}} show the probability of two-particle scattering. 

\begin{figure*}[ht!]
    \centering
    \includegraphics[scale=0.55]{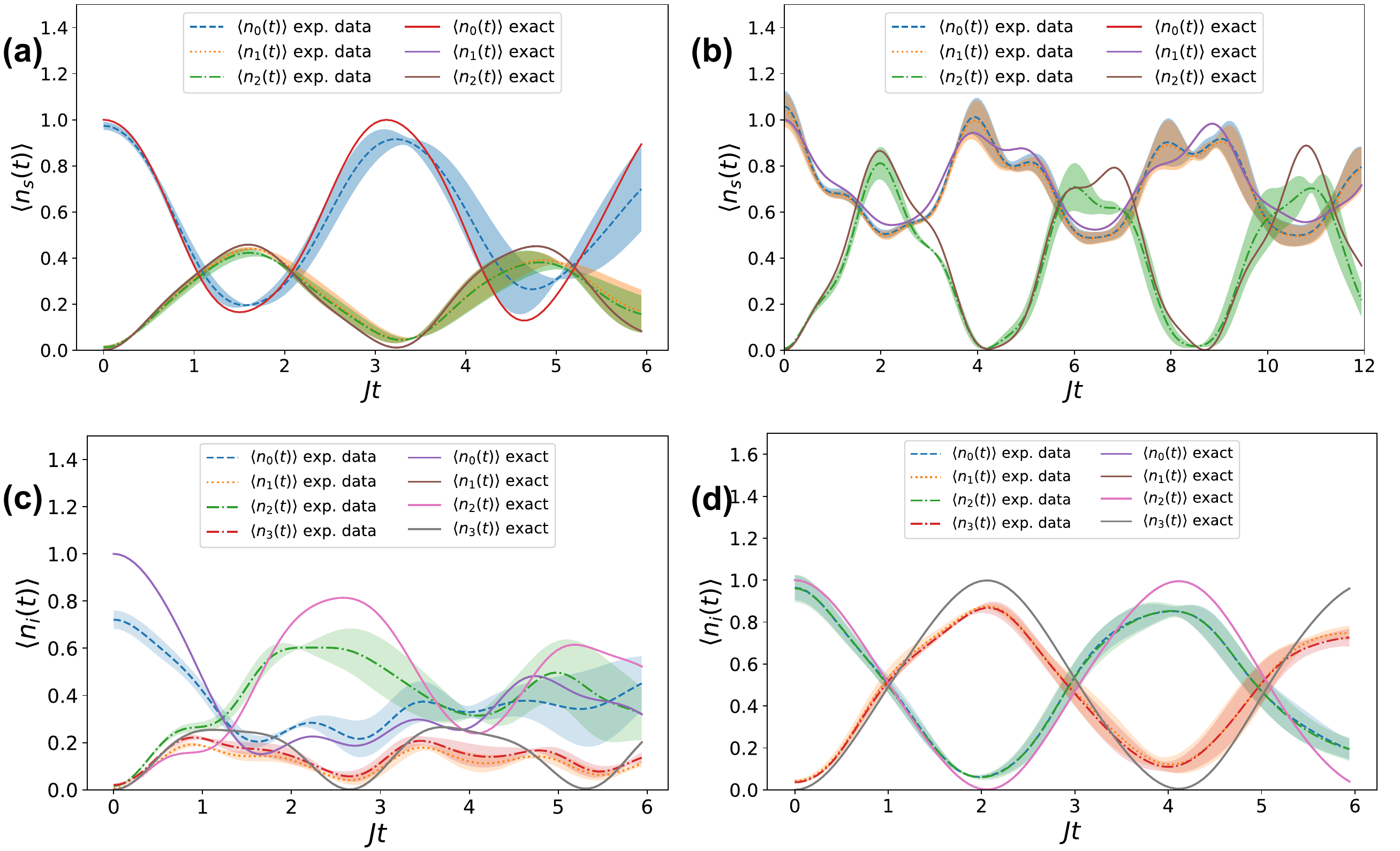}
    \caption{Occupation numbers $\langle n_i(t)$ at the $i$th spatial site \textit{vs.}~time calculated using energies obtained from exact diagonalization and compared to those calculated from ROEM energies using QLanczos algorithm on IBM Q Yorktown hardware. The initial states are {\bf{(a)}} $|100\rangle$, {\bf{(b)}} $|110\rangle$, {\bf{(c)}} $|1000\rangle$, and {\bf{(d)}} $|1010\rangle$. The parameters are set to $J=0.6$ and $h_T=1$. $N_{\text{runs}}=3$, and the shaded regions show one-standard-deviation error. In {\bf{(a)}}, {\bf{(b)}}, and {\bf{(c)}} $\langle n_1(t)\rangle$ and $\langle n_2(t) \rangle$, and in {\bf{(d)}} $\langle n_0(t)\rangle$ and $\langle n_2(t) \rangle$, as well as $\langle n_1(t)\rangle$ and $\langle n_3(t) \rangle$ overlap with each other.}
    \label{fig:n_s_average}
\end{figure*}

Similarly, in Fig.\ \ref{fig:n_s_average} 
we show a comparison between the numerical value of occupation numbers at various spatial sites calculated from a given exact $|\text{initial}\rangle$ state and the one calculated experimentally from the energy eigenvalues and corresponding eigenstates using the QLanczos algorithm. 

\begin{figure*}[ht!]
    \centering
    \includegraphics[scale=0.50]{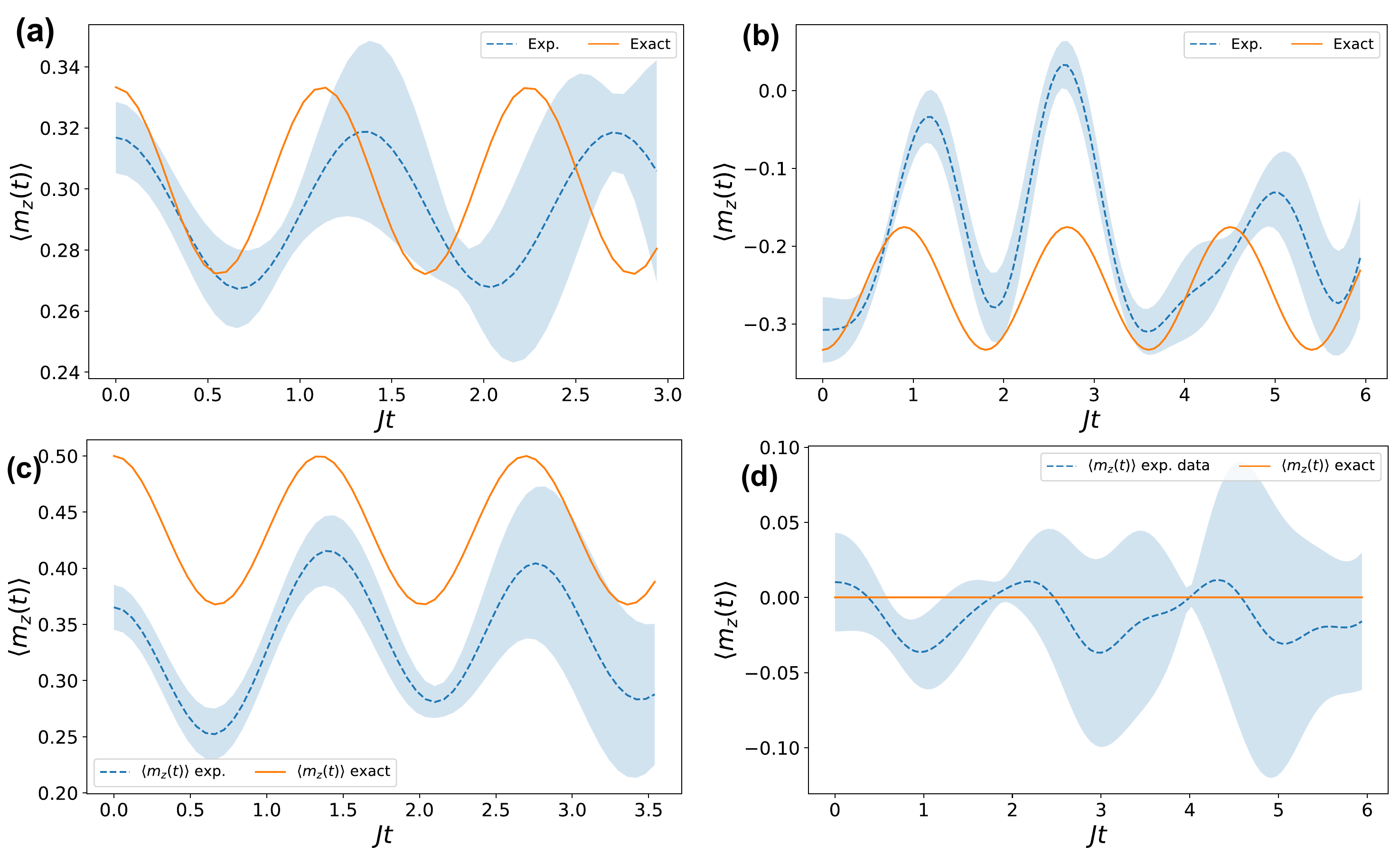}
    \caption{Exact magnetization \textit{vs.}~time calculated using energies obtained from exact diagonalization and compared to those calculated from ROEM energies using QLanczos algorithm on IBM Q Yorktown hardware. The initial states are {\bf{(a)}} $|100\rangle$, {\bf{(b)}} $|110\rangle$, {\bf{(c)}} $|1000\rangle$, and {\bf{(d)}} $|1010\rangle$. The parameters are set to $J=0.6$ and $h_T=1$. $N_{\text{runs}}=3$, and the shaded regions show one-standard-deviation error. }
    \label{fig:mz_average}
\end{figure*}

It should be noted that when the particles are initially at sites 0, 2 (i.e., $|\text{initial}\rangle=|1010\rangle$) the time evolution of the occupation number at even (odd) sites is the same, i.e., $\langle{n_0(t)}\rangle=\langle{n_2(t)}\rangle$ ($\langle{n_1(t)}\rangle=\langle{n_3(t)}\rangle$).

Finally, in Fig.\ \ref{fig:mz_average}
{\bf{(a)}}-{\bf{(d)}},
we present a comparison between the numerical value of the average transverse magnetization calculated from a given exact $|\text{initial}\rangle$ state and the experimental average transverse magnetization obtained from energy eigenvalues and corresponding eigenstates using the QLanczos algorithm.

Figs.\ \ref{fig:Transition_average}, \ref{fig:n_s_average}, and \ref{fig:mz_average} demonstrate that for number of sites $N_s=3$, the exact results and experimental data are in \fix{very good} agreement. For a larger system ($N_s=4$), the exact and experimental data are still in good agreement. In the latter case, the quantum circuit used to calculate the energy expectation values includes more single-qubit rotation and CNOT gates, which result in more error in the measurements. This can be seen by comparing Fig.\ \ref{fig:4qubit_QITE} with Fig.\ \ref{fig:3qubit_QITE} in Section~\ref{sec:QITE}.

\section{Conclusion}
\label{sec:Conclusion}

In this work, we discussed a hybrid quantum-classical method to calculate physical properties of the Ising spin chain model as a function of time, such as transition amplitudes, occupation numbers at various sites, and transverse magnetization, using the QLanczos algorithm as a tool. We took advantage of the symmetry of the system to simplify the quantum computation of the eigenvalues and eigenstates of the Hamiltonian of the system which were then used for the computation of various physical quantities of interest. We ran experiments \fix{for the QITE algorithm} on IBM Q Yorktown hardware \fix{eigenvector quantum circuits on IBM Q Vigo, Casablanca and Manhattan devices} for $N_s=3$ and $N_s=4$ spatial sites. 
Our results show good agreement with the exact values of the physical quantities of interest. It should be pointed out that although the use of the \textit{initialize} function in the IBM Qiskit library gives energy expectation value calculations at each QITE step which are very close to the exact value in the noiseless simulator case, our results show how different the noisy simulator and the hardware data can be from each other as well as exact calculations. Our data constitute the first demonstration of quantum imaginary-time evolution in a 4-qubit system on NISQ hardware, and can be useful for benchmarking purposes. 

Notably, the use of the symmetry of the system in simplifying the QITE and QLanczos algorithms reduces the number of steps in the quantum calculations, which leads to a significant reduction in error due to NISQ hardware. The QITE and QLanczos algorithms converge to the minima determined by the symmetry subgroup of the chosen initial state. Further, higher excited states can be obtained by reversing the sign of the Hamiltonian as needed. These two features enabled us to find energy levels that otherwise were difficult to compute due to the numerical difficulty associated with increasing the number of vectors in the Krylov space.

\acknowledgments
This manuscript has been authored by UT-Battelle, LLC, under Contract No. DE-AC0500OR22725 with the U.S. Department of Energy. The quantum circuits were drawn using Q-circuit package \cite{QCircuit}.
This work was supported by the Quantum Information Science Enabled Discovery (QuantISED) for High Energy Physics program at ORNL under FWP number ERKAP61 and used resources of Oak Ridge Leadership Computing Facility located at ORNL, which is supported by the Office of Science of the Department of Energy under contract No. DE-AC05-00OR22725. 
 The authors acknowledge use of the IBM Q for this work. The views expressed are those of the authors and do not reflect the official policy or position of IBM or the IBM Q team. GS is supported by ARO grant W911NF-19-1-0397 and NSF grant OMA-1937008.
\section*{Author Contributions}
K. Y. A. and G. S. designed the study, K. Y. A. collected data and produced figures. R. C. P. and G. S. supervised the research. All authors discussed the results and contributed to the final paper.

\section*{Competing Interests}
The authors declare that there are no competing interests.

\end{document}